\documentclass[useAMS,usenatbib]{mn2e}
\usepackage{multirow}
\usepackage{epsfig}
\usepackage{subfigure}

\makeatletter
\let\jnl@style=\rm
\def\ref@jnl#1{{\jnl@style#1}}

\def\aj{\ref@jnl{AJ}}                   % Astronomical Journal
\def\araa{\ref@jnl{ARA\&A}}             % Annual Review of Astron and Astrophys
\def\apj{\ref@jnl{ApJ}}                 % Astrophysical Journal
\def\apjl{\ref@jnl{ApJ}}                % Astrophysical Journal, Letters
\def\apjs{\ref@jnl{ApJS}}               % Astrophysical Journal, Supplement
\def\ao{\ref@jnl{Appl.~Opt.}}           % Applied Optics
\def\apss{\ref@jnl{Ap\&SS}}             % Astrophysics and Space Science
\def\aap{\ref@jnl{A\&A}}                % Astronomy and Astrophysics
\def\aapr{\ref@jnl{A\&A~Rev.}}          % Astronomy and Astrophysics Reviews
\def\aaps{\ref@jnl{A\&AS}}              % Astronomy and Astrophysics, Supplement
\def\azh{\ref@jnl{AZh}}                 % Astronomicheskii Zhurnal
\def\baas{\ref@jnl{BAAS}}               % Bulletin of the AAS
\def\jrasc{\ref@jnl{JRASC}}             % Journal of the RAS of Canada
\def\memras{\ref@jnl{MmRAS}}            % Memoirs of the RAS
\def\mnras{\ref@jnl{MNRAS}}             % Monthly Notices of the RAS
\def\pra{\ref@jnl{Phys.~Rev.~A}}        % Physical Review A: General Physics
\def\prb{\ref@jnl{Phys.~Rev.~B}}        % Physical Review B: Solid State
\def\prc{\ref@jnl{Phys.~Rev.~C}}        % Physical Review C
\def\prd{\ref@jnl{Phys.~Rev.~D}}        % Physical Review D
\def\pre{\ref@jnl{Phys.~Rev.~E}}        % Physical Review E
\def\prl{\ref@jnl{Phys.~Rev.~Lett.}}    % Physical Review Letters
\def\pasp{\ref@jnl{PASP}}               % Publications of the ASP
\def\pasj{\ref@jnl{PASJ}}               % Publications of the ASJ
\def\qjras{\ref@jnl{QJRAS}}             % Quarterly Journal of the RAS
\def\skytel{\ref@jnl{S\&T}}             % Sky and Telescope
\def\solphys{\ref@jnl{Sol.~Phys.}}      % Solar Physics
\def\sovast{\ref@jnl{Soviet~Ast.}}      % Soviet Astronomy
\def\ssr{\ref@jnl{Space~Sci.~Rev.}}     % Space Science Reviews
\def\zap{\ref@jnl{ZAp}}                 % Zeitschrift fuer Astrophysik
\def\nat{\ref@jnl{Nature}}              % Nature
\def\iaucirc{\ref@jnl{IAU~Circ.}}       % IAU Cirulars
\def\aplett{\ref@jnl{Astrophys.~Lett.}} % Astrophysics Letters
\def\apspr{\ref@jnl{Astrophys.~Space~Phys.~Res.}}
\def\bain{\ref@jnl{Bull.~Astron.~Inst.~Netherlands}}
\def\fcp{\ref@jnl{Fund.~Cosmic~Phys.}}  % Fundamental Cosmic Physics
\def\gca{\ref@jnl{Geochim.~Cosmochim.~Acta}}   % Geochimica Cosmochimica Acta
\def\grl{\ref@jnl{Geophys.~Res.~Lett.}} % Geophysics Research Letters
\def\jcp{\ref@jnl{J.~Chem.~Phys.}}      % Journal of Chemical Physics
\def\jgr{\ref@jnl{J.~Geophys.~Res.}}    % Journal of Geophysics Research
\def\jqsrt{\ref@jnl{J.~Quant.~Spec.~Radiat.~Transf.}}
\def\memsai{\ref@jnl{Mem.~Soc.~Astron.~Italiana}}
\def\nphysa{\ref@jnl{Nucl.~Phys.~A}}   % Nuclear Physics A
\def\physrep{\ref@jnl{Phys.~Rep.}}   % Physics Reports
\def\physscr{\ref@jnl{Phys.~Scr}}   % Physica Scripta
\def\planss{\ref@jnl{Planet.~Space~Sci.}}   % Planetary Space Science
\def\procspie{\ref@jnl{Proc.~SPIE}}   % Proceedings of the SPIE

\makeatother

\title[True Type 2 Seyfert galaxies]{Simultaneous X-ray and optical observations of true Type 2 Seyfert galaxies\thanks{Based on observations obtained with
    {\it XMM-Newton}, an ESA science mission with instruments and contributions directly funded by ESA Member States and the USA (NASA); with the TNG and NOT operated on the island
    of La Palma by the Centro Galileo Galilei and the Nordic Optical Telescope Science Association respectively, in the Spanish Observatorio del Roque de los Muchachos; at the Centro Astron\'omico Hispano Alem\'an (CAHA) at Calar Alto, operated jointly by the Max-Planck Institut f\"ur Astronomie and the Instituto de Astrof\'\i sica de Andaluc\'\i a (CSIC); at the European Organisation for Astronomical Research in the Southern Hemisphere, Chile: 278.B-5021(A), 278.B-5016(A); at the Observatorio de Sierra Nevada (OSN) operated by the Instituto de Astrof\'\i sica de Andaluc\'\i a (CSIC).}}

\author[Stefano Bianchi, et al.]{Stefano Bianchi$^1$\thanks{E-mail: bianchi@fis.uniroma3.it (SB)}, Francesca Panessa$^2$, Xavier Barcons$^3$, Francisco J. Carrera$^3$,
\newauthor Fabio La Franca$^1$, Giorgio Matt$^1$, Francesca Onori$^1$, Anna Wolter$^4$,
\newauthor Amalia Corral$^5$, Lorenzo Monaco$^6$, \'Angel Ruiz$^{3,7}$, Murray Brightman$^8$\\
$^1$Dipartimento di Fisica, Universit\`a degli Studi Roma Tre, via della Vasca Navale 84, 00146 Roma, Italy\\
$^2$Istituto di Astrofisica Spaziale e Fisica Cosmica (IASF-INAF), via del Fosso del Cavaliere 100, 00133 Roma, Italy\\
$^3$Instituto de F\'isica de Cantabria (CSIC-Universidad de Cantabria), 39005 Santander, Spain\\
$^4$INAF-Osservatorio Astronomico di Brera, via Brera 28, 20121, Milano, Italy\\
$^5$Institute of Astronomy \& Astrophysics, National Observatory of Athens, Palaia Penteli, 15236, Athens, Greece\\
$^6$European Southern Observatory, 19001 Casilla, Santiago, Chile\\
$^7$Inter-University Centre for Astronomy and Astrophysics (IUCAA), Post Bag 4, Ganeshkhind, Pune 411 007, India\\
$^8$Max-Planck-Institut f\"ur Extraterrestrische Physik, Giessenbachstrasse 1, D-85748, Garching bei München, Germany\\
}

\begin{document}

%\pagerange{\pageref{firstpage}--\pageref{lastpage}} \pubyear{2004}

\maketitle

\label{firstpage}

\begin{abstract}
We present the results of a campaign of simultaneous X-ray and optical observations of `true' Type 2 Seyfert galaxies candidates, i.e. AGN without a Broad Line Region (BLR). Out of the initial sample composed by 8 sources, one object, IC~1631, was found to be a misclassified starburst galaxy, another, Q2130-431, does show broad optical lines, while other two, IRAS~01428-0404 and NGC~4698, are very likely absorbed by Compton-thick gas along the line of sight. Therefore, these four sources are not unabsorbed Seyfert 2s as previously suggested in the literature. On the other hand, we confirm that NGC~3147, NGC~3660, and Q2131-427 belong to the class of true Type 2 Seyfert galaxies, since they do not show any evidence for a broad component of the optical lines nor for obscuration in their X-ray spectra. These three sources have low accretion rates ($\dot{m}= L_{bol}/L_{Edd} \la0.01$), in agreement with theoretical models which predict that the BLR disappears below a critical value of $L_{bol}/L_{Edd}$. The last
source, Mrk~273x, would represent an exception even of this accretion-dependent versions of the Unification Models, due to its high X-ray luminosity and accretion rate, and no evidence for obscuration. However, its optical classification as a Seyfert 2 is only based on the absence of a broad component of the H$\beta$, due to the lack of optical spectra encompassing the H$\alpha$ band.
\end{abstract}


\section{Introduction}

The fundamental idea behind the standard Unified Model \citep{antonucci93} is that type 1 and type 2 Active Galactic Nuclei (AGN) have no intrinsic physical differences, their classification being instead determined by the presence or not of absorbing material along the line-of-sight to the object. This scenario has been extremely successful, although some additional ingredients are needed in order to take into account all the observational evidence \citep[see e.g.][for a review]{bmr12}. Among the failed expectations of the Unification Model is the lack of broad optical lines in the polarized spectra of about half of the brightest Seyfert 2 galaxies, even when high-quality spectropolarimetric data are available \citep[e.g.][]{tran01,tran03}.

These Seyfert 2 galaxies without an hidden Broad Line Region (BLR) are observationally found to accrete at low Eddington rates \citep[e.g.][]{nic03,bigu07,shu07,wu11,mar12}. This is in agreement with theoretical models which predict that the BLR disappears below a certain critical value of accretion rate and/or luminosity \citep[e.g.][]{nic00,eh09,trump11}. If the BLR cannot form in weakly accreting AGN, we expect the existence of the unobscured counterparts of the non-hidden BLR Seyfert 2 galaxies, that is, optically classified Type 2 objects, without any evidence of obscuration in their X-ray spectrum. In the last ten years, a significant number of such `true' Type 2 Seyfert galaxies have been claimed in the literature \citep[e.g.][]{pappa01,pb02,boll03,wol05,gliozzi07,bianchi08a,brna08,panessa09,shi10,tran11,trump11}, with one case in which the BLR is present but characterised by an intrinsically high Balmer decrement \citep{corr05}. Most of these sources have the low accretion rates/luminosities required
by the above-mentioned theoretical models.

Nevertheless, an homogeneous and unambiguous sample of true Type 2 Seyfert galaxies is still missing. One of the main issues is that sources may be highly variable and may change their optical and/or X-ray appearance in different observations. These ‘changing-look’ AGN are not uncommon. In some cases, this behaviour is best explained by a real `switching-off' of the nucleus \citep[see e.g.][]{gilli00,gua05}, but a variable column density of the absorber appears as the best explanation in the majority of the cases \citep[e.g.][and references therein]{elvis04,ris05,bianchi09c,bmr12}. If the optical and the X-ray spectrum are taken in two different states of the source, it is clear that the disagreement between the two classifications may be only apparent. Therefore, the key to find genuine ‘unabsorbed Seyfert 2s’ is to perform simultaneous X-ray and optical observations. Only two unabsorbed Seyfert 2 candidates had been observed so far simultaneously in the X-rays and in the optical band, leading to the
discovery of two unambiguous true type Seyfert 2s: NGC~3147 \citep{bianchi08a} and Q2131-427 \citep{panessa09}. In this paper, we report on the results of the complete systematic campaign of simultaneous X-ray and optical observations of 8 true Type 2 Seyfert galaxies candidates.

\section{The sample}

In order to build a comprehensive sample of unabsorbed Seyfert 2 candidates, we started from the \citet{pb02} sample, which includes 17 type 2 Seyfert galaxies with an X-ray column density lower than $10^{22}$ cm$^{-2}$ and very unlikely to be Compton-thick, as suggested by isotropic indicators. We selected a conservative subsample, excluding the sources where an intrinsic column density (even if much lower than the one expected from the optical properties) is actually measured. Therefore, we only choose sources which are genuinely unabsorbed, in the sense that no column density in excess of the
Galactic one has ever been observed in the X-rays, with tight upper limits \citep[at most few $10^{21}$ cm$^{-2}$:][]{pb02}. Moreover, three other sources (NGC~4565, NGC~4579 and IRAS 20051-1117) were excluded because were found to be misclassified in the optical, all having broad components of the emission lines \citep{ho97,georg04}. Another one (NGC~7590) was found to be a Compton-thick Seyfert 2 dominated by a nearby off-nuclear ultra-luminous X-ray source \citep{shu10}, while NGC~7679 is dominated by starburst emission in the optical band \citep{dc01}.

To the remaining five objects, we added two sources belonging to a sample of `naked' AGN, i.e. spectroscopically classified as Seyfert 2s, but with very large amplitude variations in the $B_J$ passband, typical of type 1 objects, where the nucleus is directly seen without intervening absorption \citep{hawk04}. The sources included in our sample were selected from the three objects with \textit{Chandra} data presented by \citet{gliozzi07}, supporting their unabsorbed nature, excluding Q2122-444, which, re-observed simultaneously with XMM-\textit{Newton} and NTT, revealed the presence of broad optical line components, thus ruling out the true type 2 hypothesis \citep{gliozzi10}. Finally, we added to our sample NGC~3660, a very promising unabsorbed Seyfert 2 candidate suggested by \citet{brna08}.

The final sample (Table \ref{sample}) is constituted by 8 sources, which we observed with XMM-\textit{Newton}. As mentioned in the Introduction, to avoid any possible misclassification due to variability of the sources, we coordinated all the XMM-\textit{Newton} observations with quasi-simultaneous ground-based optical spectroscopy. As a final note, we would like to stress that, given the heterogeneous selection methods described above, this sample is by no means complete in any sense.

\begin{table*}\scriptsize
\begin{center}
\caption{\label{sample}The sample of true type Seyfert 2 candidates analysed in this paper, along with the details of the quasi-simultaneous X-ray and optical/NIR observations.}
\begin{tabular}{lcccccccccc}
Name  &   z   & \multicolumn{3}{c}{XMM-{Newton}} & \multicolumn{6}{c}{Optical/NIR}\\
& & Obs. Date & \textsc{obsid} & Exp. & Obs. Date & Instr. & Slit & Exp. & Range & $\lambda/\Delta\lambda$\\
\hline
& & && &&&&&&\\
IC~1631 & 0.030968 & 2006-11-23 & 0405020801 & 22 & 2006-12-01 & VLT/FORS2/300V & 1.0\arcsec & 1200 &4450-8700 & 440\\[0.2cm]
IRAS~01428-0404 & 0.018199 & 2008-08-03 & 0550940201 & 18 & 2008-07-30 & CAHA/CAFOS/G-100 & 1.8\arcsec & 1800x2 & 4900-7800 & 880\\[0.2cm]
\multirow{4}*{Mrk~273x} & \multirow{4}*{0.458000} & 2010-05-13 & 0651360301 & 14 &
\multirow{4}*{2010-05-17} & \multirow{4}*{TNG/LRS/LR-R} & \multirow{4}*{1.0\arcsec} & \multirow{4}*{1800} & \multirow{4}*{4470-10073} & \multirow{4}*{714}\\
& & 2010-05-15 & 0651360501 & 15 & & &&&&\\
& & 2010-05-17 & 0651360601 & 14 & & &&&&\\
& & 2010-06-26 & 0651360701 & 11 & & &&&&\\[0.2cm]
\multirow{3}*{NGC~3147} & \multirow{3}*{0.009346} & \multirow{3}*{2006-10-06} &
\multirow{3}*{0405020601} & \multirow{3}*{17} & 2006-10-04 & \multirow{3}*{OSN/Albireo} & \multirow{3}*{2.0\arcsec} & \multirow{3}*{1800x6} & \multirow{3}*{4000-7000} & \multirow{3}*{1500}\\
& & &&& 2006-10-05&&&&&\\
& & &&& 2006-10-09&&&&&\\[0.2cm]
\multirow{5}*{NGC~3660} & \multirow{5}*{0.012285} & \multirow{5}*{2009-06-03} &
\multirow{5}*{0601560201} & \multirow{5}*{15} & 2010-01-15 & NOT/ALFOSC/Gr7 & 0.5\arcsec & 300x2 & 3850-6850 & 1300\\
& & &&& 2011-02-04 & TNG/LRS/LR-B & 0.7\arcsec & 120 & 3000-8430 & 835\\
& & &&& 2011-02-04 & TNG/LRS/LR-R & 0.7\arcsec & 60 & 4470-10073 & 1020\\
& & &&& 2011-00-00 & TNG/NICS/IJ & 0.75\arcsec & 360x4 &9000-14500 & 665 \\
& & &&& 2011-00-00 & TNG/NICS/JS & 0.75\arcsec & 945x4 &11700-13300 & 1600\\[0.2cm]
NGC~4698 & 0.003366 & 2010-06-09 & 0651360401 & 33 & 2010-12-28 & NOT/ALFOSC/Gr7 & 0.5\arcsec & 30x2 & 3850-6850 & 1300\\[0.2cm]
Q2130-431 & 0.266 & 2006-11-13 & 0402460201 & 32 & 2006-12-20 & NTT/EMMI/Gr4 & 1.0\arcsec & 1200 & 5500-10000 & 613 \\[0.2cm]
Q2131-427 & 0.365 & 2006-11-15 & 0402460401 & 27 & 2006-12-20 & NTT/EMMI/Gr4 & 1.0\arcsec & 1200 & 5500-10000 & 613 \\[0.2cm]
\end{tabular}
\end{center}

Notes: \textit{Col (1)}: Source; \textit{Col. (2)}: redshift (NED);
\textit{Col (3-5)}: XMM-\textit{Newton} observation date, \textsc{obsid} and exposure time (ks); \textit{Col. (6-11)}: Optical/NIR observation date, instruments, adopted slit, exposures (s), wavelength range (\AA) and resolution.
\end{table*}

\section{Observations and data reduction}

\subsection{\label{xmmred}\textit{XMM-Newton}}

The XMM-\textit{Newton} observations of the sources of our sample are listed in
Table~\ref{sample}. In all cases, the observations were performed with the
EPIC CCD cameras, the pn and the two MOS, operated in Large and Small Window, respectively, and Medium Filter. Data
were reduced with SAS 8.0.0 and screening for intervals of flaring particle
background was done consistently
with the choice of extraction radii, in an iterative process based on the
procedure to maximize the signal-to-noise ratio described in detail by \citet{pico04} in their Appendix A.
The background spectra were extracted from
source-free
circular regions with a radius of 50 arcsec. The MOS data have been only used
when the number of counts in their spectra was high enough to significantly help
in the analysis. Finally, spectra were binned in order to oversample the
instrumental resolution by at least a factor of 3 and to have no less than 20
counts in each
background-subtracted spectral channel. The latter requirement allows us to use
the $\chi^2$ statistics. When the number of counts is too low (as in the cases of IRAS~01428-0404 and NGC~4698), we performed a
binning with less counts per bin, and adopted the \citet{cash76} statistics.
Similarly, in order to look for the iron K$\alpha$ emission lines in low-counts spectra, we also
performed local fits with the unbinned spectra (in a restricted energy band of 350 channels centered around rest-frame 6.4
keV) with the Cash statistics \citep[see e.g.][]{gua05b,bianchi06}.

In the following, errors and upper limits correspond to the 90 per cent confidence level for one interesting parameter ($\Delta \chi^2 =2.71$), where not otherwise stated. The adopted cosmological parameters are $H_0=70$ km s$^{-1}$ Mpc$^{-1}$ , $\Omega_\Lambda=0.73$ and $\Omega_m=0.27$ \citep[i.e. the default ones in \textsc{xspec 12.7.1}:][]{xspec}.

\subsection{Optical data}

Optical spectroscopy of our `true' Seyfert 2 candidates was obtained in a variety of ground-based telescopes and instruments, as detailed in Table~\ref{sample}. All the optical observations took place within a few days up to few months of the XMM-\textit{Newton} X-ray observations. For the purposes of our study, these observations can be considered `simultaneous', since we expect the optical spectroscopy and the X-ray observations to map essentially the sources in the same configuration.

X-ray absorption variability has been observed on a large number of sources in time-scales as short as less than a day, including temporary `eclipses' of otherwise unobscured objects \citep[e.g][]{elvis04,ris05,elvis04,puc07,bianchi09c,ris11}. However, absorption on these scales, being well within the sublimation radius and hence dust-free, cannot affect the reddening of the BLR. Therefore, these short-term variability could only explain the observation of an X-ray obscured Seyfert 1 galaxy (if observed when the cloud absorbs the X-ray source), but not the X-ray unabsorbed Seyfert 2 galaxies we are interested in.

On the other hand, the BLR can only be reddened by an absorber at a distance greater than the dust sublimation radius, which can be roughly written as $r_d\simeq0.04\,L_{43}^{1/2}$ pc, with $L_{43}$ being the bolometric luminosity in $10^{43}$ erg s$^{-1}$ \citep[adapted from][]{barv87}. In order to cover the entire BLR, a cloud must have at least the same dimensions, which again can be roughly expressed as $r_b\simeq0.008\,L_{43}^{1/2}$ pc \citep[adapted from the relationship for the H$\beta$ line presented in][]{bentz09}. The minimum time $t_m$ needed in order to completely cover or uncover the BLR is, therefore, the crossing time of such a cloud: $v=r_b/t_m$. Assuming that the cloud is in Keplerian motion around the central BH at distance $r_d$, we find an estimate of $t_m$ as

\begin{equation}
t_m = 7.9\times10^7\,L_{43}^{3/4}\,M_{8}^{-1/2}\,\mathrm{s}
\end{equation}

Therefore, even the $\simeq6$ months delay between the X-ray and optical observation in NGC~3660 is safely shorter than the minimum variability time-scale for a reddening change of the BLR estimated for this source, which is few years from the above formula and the BH mass and bolometric luminosity reported in Sect.~\ref{truetype2}.

The optical long-slit spectrographs used had a variety of spectral dispersions, typically dubbed `intermediate', i.e. enough to measure the width of an emission line of several 100 km s$^{-1}$ intrinsic width.  Observations were done with the slit oriented in parallactic angle in order not to loose flux at bluer wavelengths. The data were reduced using standard processing techniques, including de-biassing, flat-fielding along the spectral direction, arc lamp wavelength calibration, spectral extraction and background subtraction from the nearby sky and approximate flux rectification using spectrophotometric standard stars (but not correction for slit width). Statistical errors were properly propagated through the whole process, enabling us to perform statistical modelling of patches of the spectra, which we used to deblend and characterize emission lines. We typically took spectral regions around the H$\beta\lambda4861$ and $\mathrm{[{O\,\textsc{iii}}]}\lambda5007$ emission lines and the H$\alpha\lambda6563$
and $\mathrm{[{N\,\textsc{ii}}]}\lambda6583$ lines. Spectra were fitted, via $\chi^2$ minimisation, by modelling the continuum as a powerlaw or a \textsc{spline} function, and each line component as a Gaussian. In order to disentangle any broad component of the H$\alpha$, we assumed that (i) F([{N\,\textsc{II}]} $\lambda$6583)/F([{N\,\textsc{II}] $\lambda$6548) = 3, as required by the ratio of the respective Einstein coefficients, (ii) $\lambda_2$/$\lambda_1$ = 6583.39/6548.06 and (iii) the [{N\,\textsc{II}] lines are Gaussians with the same width. With the exception above, the central wavelengths of all lines
were left free, not constraining them to have a fixed ratio between them or to be at the expected redshift. The line flux ratios were then plotted in the diagrams shown in Fig.~\ref{diagn}, where the separations between Starburst galaxies, Seyfert galaxies and LINERs are marked as in \citet{kew06}.

\section{Spectral analysis}

\begin{table}
\caption{\label{tableopt}Optical line diagnostics from the analysis of the optical spectra of our campaign (ratios are expressed in
decadic logarithms). Classifications (SB: Starburst; S1: Seyfert 1; S2: Seyfert 2) are done after \citet{kew06}, and references
therein (see Fig.~\ref{diagn}).}
\begin{center}
% use packages: array
\begin{tabular}{llllll}
Source & $\frac{\mathrm{[{N\,\textsc{ii}}]}}{\mathrm{H\alpha}}$ &
$\frac{\mathrm{[{O\,\textsc{iii}}]}}{\mathrm{H\beta}}$ &
$\frac{\mathrm{[{S\,\textsc{ii}}]}}{\mathrm{H\alpha}}$  &
$\frac{\mathrm{[{O\,\textsc{i}}]}}{\mathrm{H\alpha}}$ & Cl.\\
\hline
IC~1631 & -0.58 & 0.08 & -0.63 & -1.53 & SB \\
IRAS~01428-0404 & -0.20 & 0.71 & -0.57 & -1.08 & S2\\
Mrk~273x & -- & 0.85 & -- & -- & S2? \\
NGC~3147 & 0.3 & 0.85 & 0.2 & -- & S2 \\
NGC~3660 (NOT) & -0.21 & 0.48 & -0.55 & -1.49 & S2 \\
NGC~3660 (TNG) & -0.16 & 0.63 & -0.60 & -1.43 & S2 \\
NGC~4698 & 0.42 & $>0.93$ & -0.10 & $<-0.54$ & S2 \\
Q2130-431 & 0.23 & 1.03 & -0.29 & -- & S1 \\
Q2131-427 & 0.12 & 0.97 & -- & -- & S2 \\
\hline
\end{tabular}
\end{center}
\end{table}

\begin{figure*}
\begin{center}
\epsfig{file=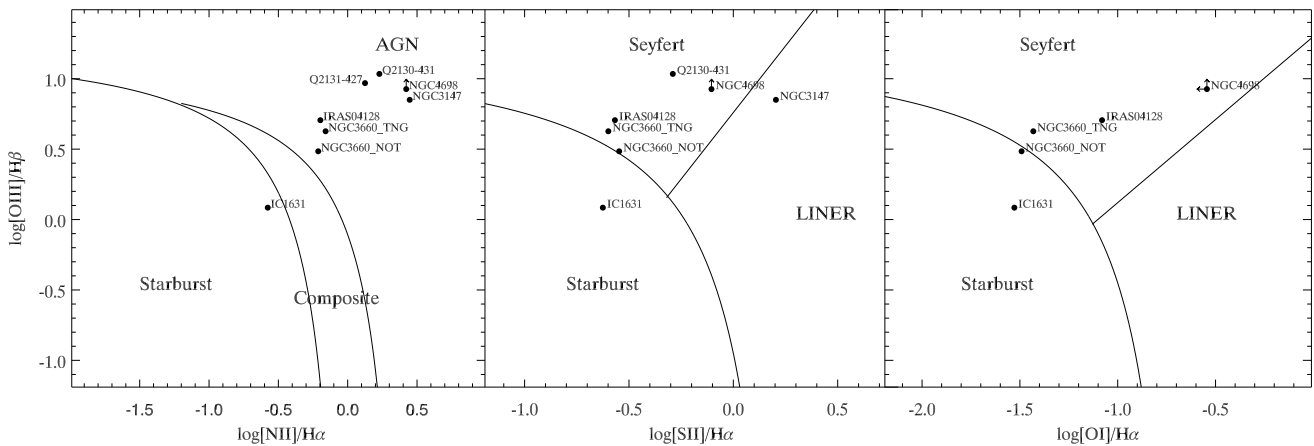,width=\textwidth}
\end{center}
\caption{\label{diagn}Optical line diagnostic diagrams from the analysis of the optical spectra of our campaign. Arrows indicate upper/lower limits. Classifications are done after \citet{kew06}, and references therein (see Table~\ref{tableopt}).}
\end{figure*}

\subsection{IC1631}

The spectral classification of this source as given by \citet{pb02} was rather
ambiguous, because of the lack of a measured flux of $\mathrm{[{O\,\textsc{i}}]}6300$.
Indeed, their optical data was taken from \citet{sw93}, who classified IC~1631
as a starburst galaxy. On the other hand, \citet{ks90} opted for an AGN
classification, based on the large FWHM of the $\mathrm{[{O\,\textsc{iii}}]}$
line. The optical line ratios derived from our high-quality spectrum, shown in
Fig~\ref{ic1631_optical}}, all clearly points to a classification as a starburst galaxy (see Tables~\ref{tableopt} and
\ref{ic1631_eso}, and Fig.~\ref{diagn}), with no sign of AGN activity in this
galaxy.

\begin{table}
\caption{\label{ic1631_eso}IC1631: optical emission lines in the \textit{ESO} spectrum (see Fig.~\ref{ic1631_optical}).}
\begin{center}
% use packages: array
\begin{tabular}{llll}
Line & $\lambda_\mathrm{l}$ & FWHM & Flux \\
(1) & (2) & (3) & (4)\\
\hline
 &  &  &\\
H$\beta$ & 4861.33 & $490\pm50$ & $2.8\pm0.3$ \\[1ex]
$\mathrm{[{O\,\textsc{iii}}]}$ & 4958.92 & $540\pm60$ &
$1.0\pm0.2$\\[1ex]
$\mathrm{[{O\,\textsc{iii}}]}$ &  5006.85 & $540\pm60$ &
$3.4\pm0.4$\\[1ex]
$\mathrm{[{O\,\textsc{i}}]}$ & 6300.32 & $1000\pm700$ &
$0.4\pm0.3$\\[1ex]
$\mathrm{[{N\,\textsc{ii}}]}$ & 6548.06 & $415\pm40$ &
$1.19\pm0.03$\\[1ex]
H$\alpha$ & 6562.79 & $414\pm18$ & $13.5\pm0.6$ \\[1ex]
$\mathrm{[{N\,\textsc{ii}}]}$ & 6583.39 & $415\pm40$ &
$3.58\pm0.09$\\[1ex]
$\mathrm{[{S\,\textsc{ii}}]}$ & 6716.42 & $430\pm50$ &
$1.8\pm0.3$\\[1ex]
$\mathrm{[{S\,\textsc{ii}}]}$ & 6730.78 & $430\pm50$ &
$1.4\pm0.3$\\
 &  &  &  \\
\hline
\end{tabular}
\end{center}
\textsc{Notes.}-- Col. (1) Identification. Col (2) Laboratory wavelength (\AA),
in air \citep{bowen60}. Col. (3) km s$^{-1}$ (instrumental resolution not removed).
 Col. (4) $10^{-14}$ erg cm$^{-2}$ s$^{-1}$.

\end{table}

\begin{figure*}
\begin{center}
\epsfig{file=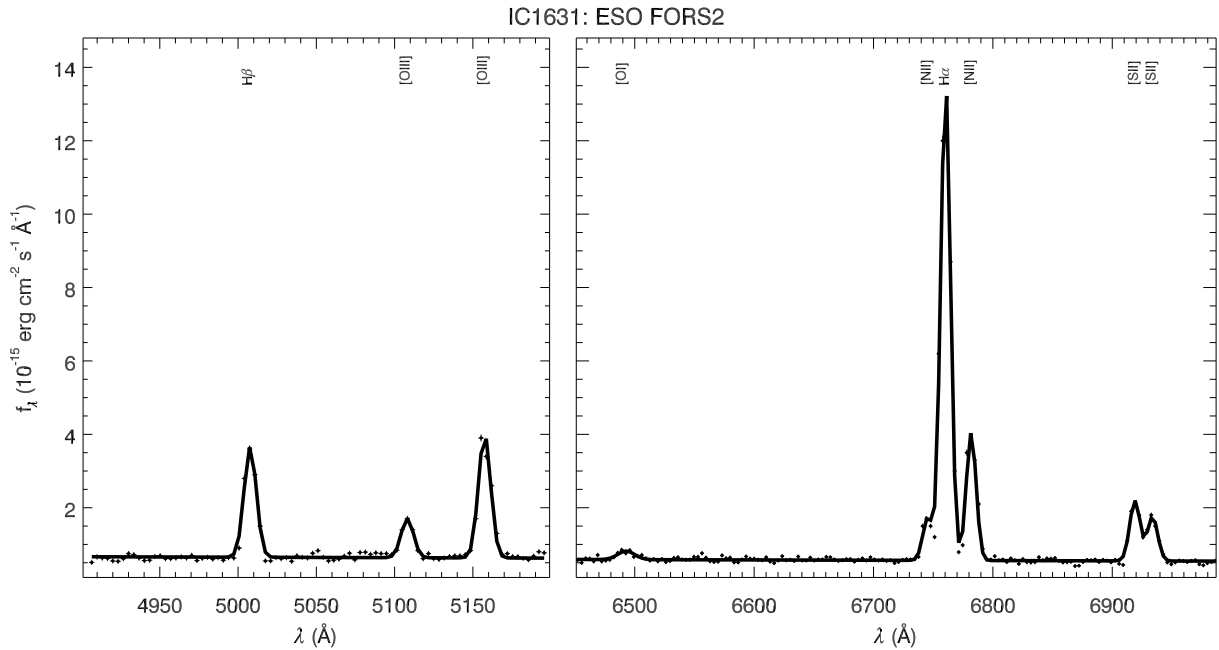,width=\textwidth}
\end{center}
\caption{\label{ic1631_optical}IC~1631: ESO optical spectrum and best fit (see Table~\ref{ic1631_eso}).}
\end{figure*}

In X-rays, IC~1631 had never been observed before our campaign, with the
only exception of a non-detection by \textit{Ginga}, with a very loose upper
limit for the 2-10 keV flux of $1\times10^{-11}$ erg cm$^2$ s$^{-1}$, as it
was likely contaminated by Abell 2877 \citep{awaki93}. Although the source is
quite faint, the XMM-Newton EPIC pn image appears extended, with roughly an excess in counts of 40-80\% with respect to the PSF between 10 and 15\arcsec, and no variability
is observed. The X-ray spectrum can be fitted ($\chi^2=14/13$ d.o.f.) with a simple powerlaw ($\Gamma=2.0\pm0.3$),
absorbed only by the Galactic column density (Fig.~\ref{ic1631_xmm}). Any absorption in excess, at the redshift of the source,
can be constrained to be lower than $4\times10^{20}$ cm$^{-2}$. No iron line is
required by the data, with an upper limit to its flux of $2\times10^{-7}$ ph
cm$^{-2}$ s$^{-1}$ in a local fit to the unbinned spectrum (due to the low level
of continuum at this energy, no upper limit to the EW can be estimated). More refined models cannot be tested due to the low quality of the spectrum. However, the addition of a thermal component (\textsc{apec} in \textsc{xspec}) gives a very good fit ($\chi^2=7/11$ d.o.f.) with kT=$0.3^{+0.2}_{-0.1}$ keV and $\Gamma=1.5\pm0.3$, in agreement with the expectations for starburst galaxies \citep[e.g.][]{pr02}.

The 2-10 (0.5-2) keV flux is $3.0\pm1.0$ ($2.3\pm0.3$) $\times10^{-14}$ erg
cm$^{-2}$ s$^{-1}$, corresponding to a 2-10 keV luminosity of
$6.6\pm2.1\times10^{40}$ erg s$^{-1}$. This luminosity would correspond to a
star formation rate of $\simeq10-15$ M$_\odot$/yr, according to the relations
presented in \citet{ranalli03}, which is well in the range of local starburst galaxies \citep[e.g.][]{sw09}.

\begin{figure}
\begin{center}
\epsfig{file=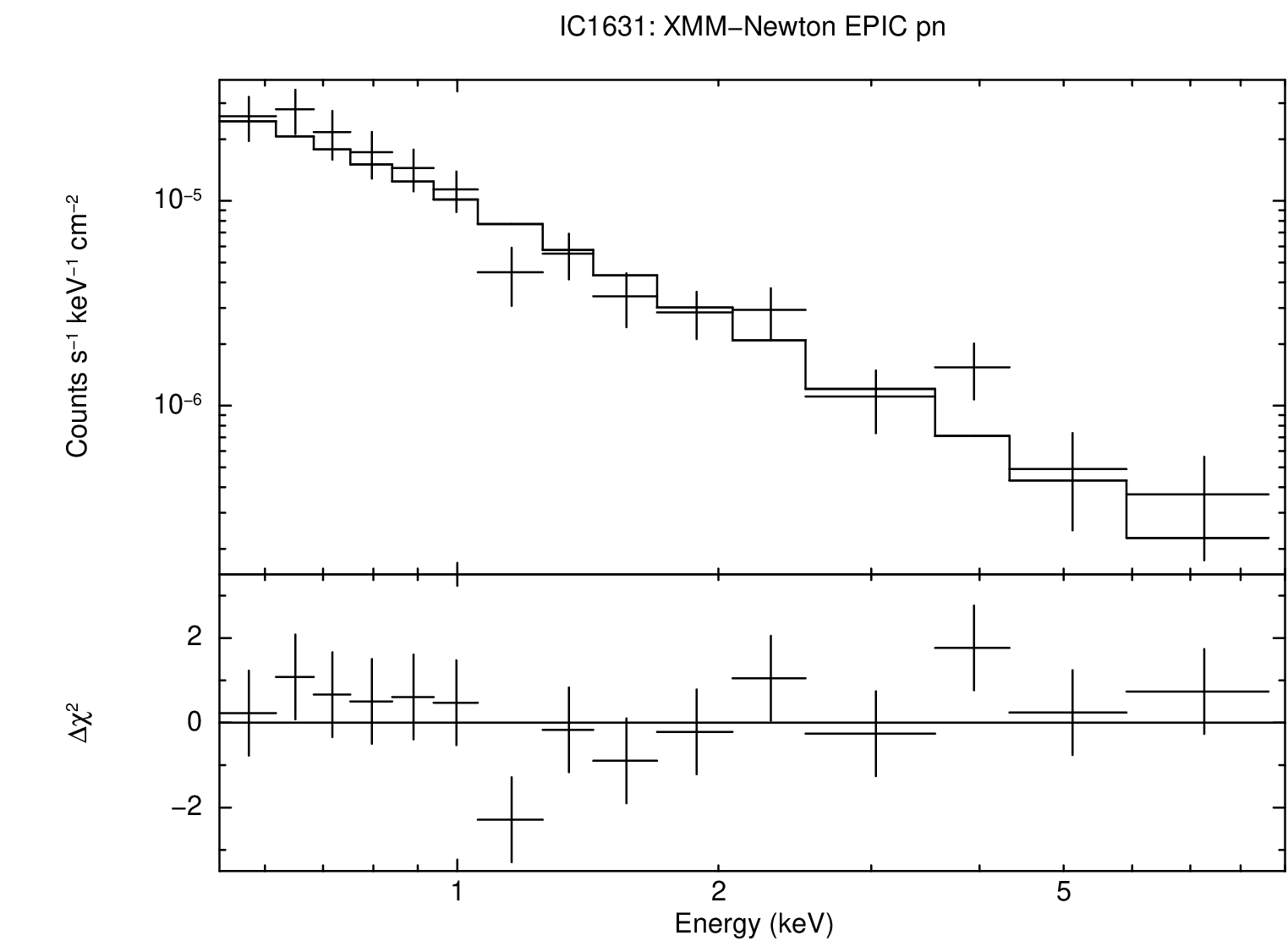,angle=-90,width=0.45\textwidth}
\end{center}
\caption{\label{ic1631_xmm}IC~1631: the XMM-\textit{Newton} observation.}
\textit{EPIC pn spectrum, along with the best fit and residuals in terms of $\Delta\chi^2$. See text for details.}
\end{figure}

\subsection{IRAS01428-0404}

IRAS~01428-0404 was optically classified as Seyfert 2 by \citet{moran96} and
\citet{pie98}. In our new optical spectrum (Fig.~\ref{iras_optical}), the line ratios all clearly confirm this classification (see Tables~\ref{tableopt}, \ref{iras_caha}, and Fig.~\ref{diagn}).

\begin{table}
\caption{\label{iras_caha}IRAS01428-0404: optical emission lines in the CAHA spectrum (see Fig~\ref{iras_optical}).}
\begin{center}
% use packages: array
\begin{tabular}{llll}
Line & $\lambda_\mathrm{l}$ & FWHM & Flux \\
(1) & (2) & (3) & (4)\\
\hline
 &  &  & \\
H$\beta$ & 4861.33 & $275\pm15$ & $5.2\pm0.3$ \\[1ex]
$\mathrm{[{O\,\textsc{iii}}]}$ & 4958.92 & $318\pm4$ &
$8.7\pm0.3$\\[1ex]
$\mathrm{[{O\,\textsc{iii}}]}$ & 5006.85 & $318\pm4$ &
$26.4\pm0.4$\\[1ex]
$\mathrm{[{O\,\textsc{i}}]}$ & 6300.32 & $400\pm30$ &
$3.9\pm0.3$\\[1ex]
$\mathrm{[{N\,\textsc{ii}}]}$ & 6548.06 & $360\pm4$ &
$9.92\pm0.12$\\[1ex]
H$\alpha$ & 6562.79 & $364\pm3$ & $46.9\pm0.4$ \\[1ex]
$\mathrm{[{N\,\textsc{ii}}]}$ & 6583.39 & $360\pm4$ &
$29.8\pm0.4$\\[1ex]
$\mathrm{[{S\,\textsc{ii}}]}$ & 6716.42 & $345\pm11$ &
$6.6\pm0.4$\\[1ex]
$\mathrm{[{S\,\textsc{ii}}]}$ & 6730.78 & $345\pm11$ &
$6.1\pm0.3$\\
 &  &  & \\
\hline
\end{tabular}
\end{center}
\textsc{Notes.}-- Col. (1) Identification. Col (2) Laboratory wavelength (\AA),
in air \citep{bowen60}. Col. (3) km s$^{-1}$ (instrumental resolution not removed).
 Col. (4) $10^{-16}$ erg cm$^{-2}$ s$^{-1}$.

\end{table}

\begin{figure*}
\begin{center}
\epsfig{file=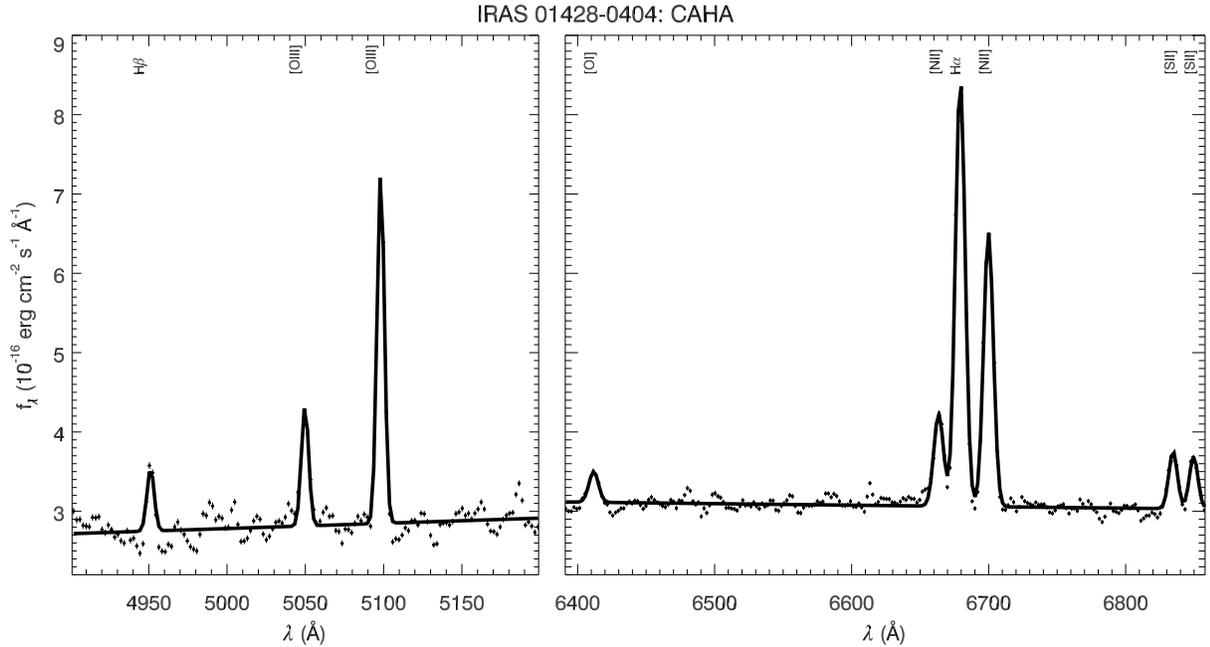,width=\textwidth}
\end{center}
\caption{\label{iras_optical}IRAS01428-0404: CAHA optical spectrum and best fit (see Table~\ref{iras_caha}).}
\end{figure*}

In X-rays, IRAS~01428-0404 was detected in the \textit{ROSAT} All-Sky Survey (performed in the second half of 1990) with a 0.1-2.4 keV flux of $6.3\times10^{-12}$ erg s$^{-1}$ cm$^{-2}$ \citep{boller92}, but the identification was subsequently dubbed as insecure by \citet{boll98}, and by \textit{ASCA} (in 1997) with a reported 2-10 keV flux of $4\times10^{-13}$ erg s$^{-1}$ cm$^{-2}$ \citep{pb02}. As for our XMM-\textit{Newton} observation, the simplest possible model, i.e. a power law
absorbed by the Galactic column density, provides a good fit (normalized Cash statistics is 39/36 d.o.f.) to the 0.5-8 keV spectrum (the source is not detected above this energy: see Fig.~\ref{iras01428_xmm}). The powerlaw index is $\Gamma=2.3\pm0.3$, and no intrinsic absorption is found at the redshift of the source
(N$_H<6\times10^{20}$ cm$^{-2}$). No significant variability is observed during the observation. The 2-10 (0.5-2) keV observed flux is $2.1\pm0.5$ ($3.0\pm0.5$)$\times$10$^{-14}$ erg s$^{-1}$ cm$^{-2}$, corresponding to an
absorption-corrected luminosity of $1.9\pm0.4$ ($2.5\pm0.4$)$\times$10$^{40}$ erg s$^{-1}$.  The 2-10 keV to $\mathrm{[{O\,\textsc{iii}}]}$ luminosity ratio,
once corrected for the Balmer decrement as measured in our optical spectrum, is $\log{\frac{L_X}{L_{[OIII]}}}\simeq-0.5$, strongly suggesting that the source is
Compton-thick \citep[see e.g.][]{panessa06,lamastra09,mar12}. Although flatter powerlaw indices are expected for X-ray spectra of Compton-thick sources, much steeper ones are usually found in low-counts sources, since the dominating component is the soft excess which peaks where the instrumental effective area is larger, instead of the Compton reflection component at higher energies \citep[see e.g.][]{bianchi06}.

This interpretation is also in agreement with the overall Spectral Energy Distribution (SED), shown in the left panel of
Fig.~\ref{irassed}, which resembles that of a Compton-thick source.  The SED has been built
using data from NED, SuperCosmos Sky Survey (SSS) and our XMM-Newton proprietary
data (photometric data from the Optical Monitor and re-binned X-ray spectrum).
The SED is corrected for Galactic absorption and shifted to rest frame. We
fitted this SED with a set of pure AGN and pure SB templates \citep[see][for a
complete description of these templates]{ruiz10}, finding as best fit model a
combination of a type 2 AGN and a starburst. The AGN template (NGC 3393) has an
Hydrogen column density $>10^{25}$ cm$^{-2}$. The starburst component (template
from IRAS 12112+0305) is $\simeq40\%$ of the total bolometric output of this
source.

\begin{figure}
\begin{center}
\epsfig{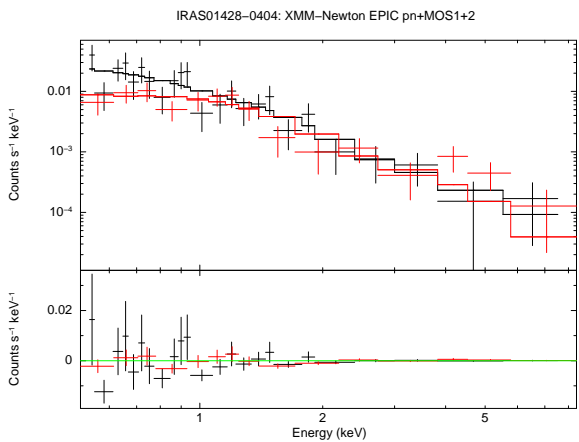}
\end{center}
\caption{\label{iras01428_xmm}IRAS~01428-0404: the XMM-\textit{Newton} observation. EPIC pn (black) and co-added MOS (red) spectra, along with the best fit and residuals. See text for details.}
\end{figure}

\begin{figure*}
\begin{center}
\epsfig{file=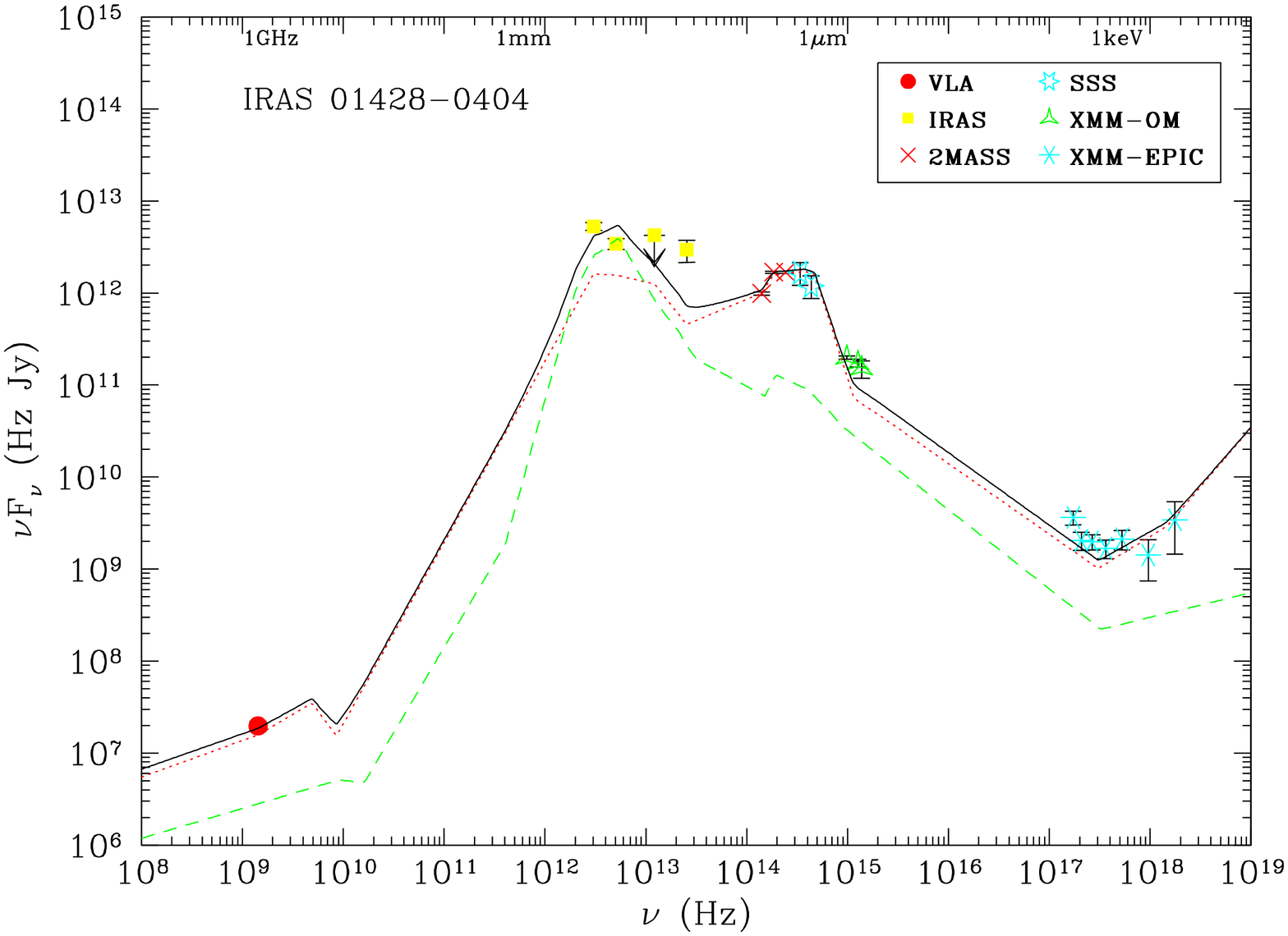,width=0.45\textwidth}
\epsfig{file=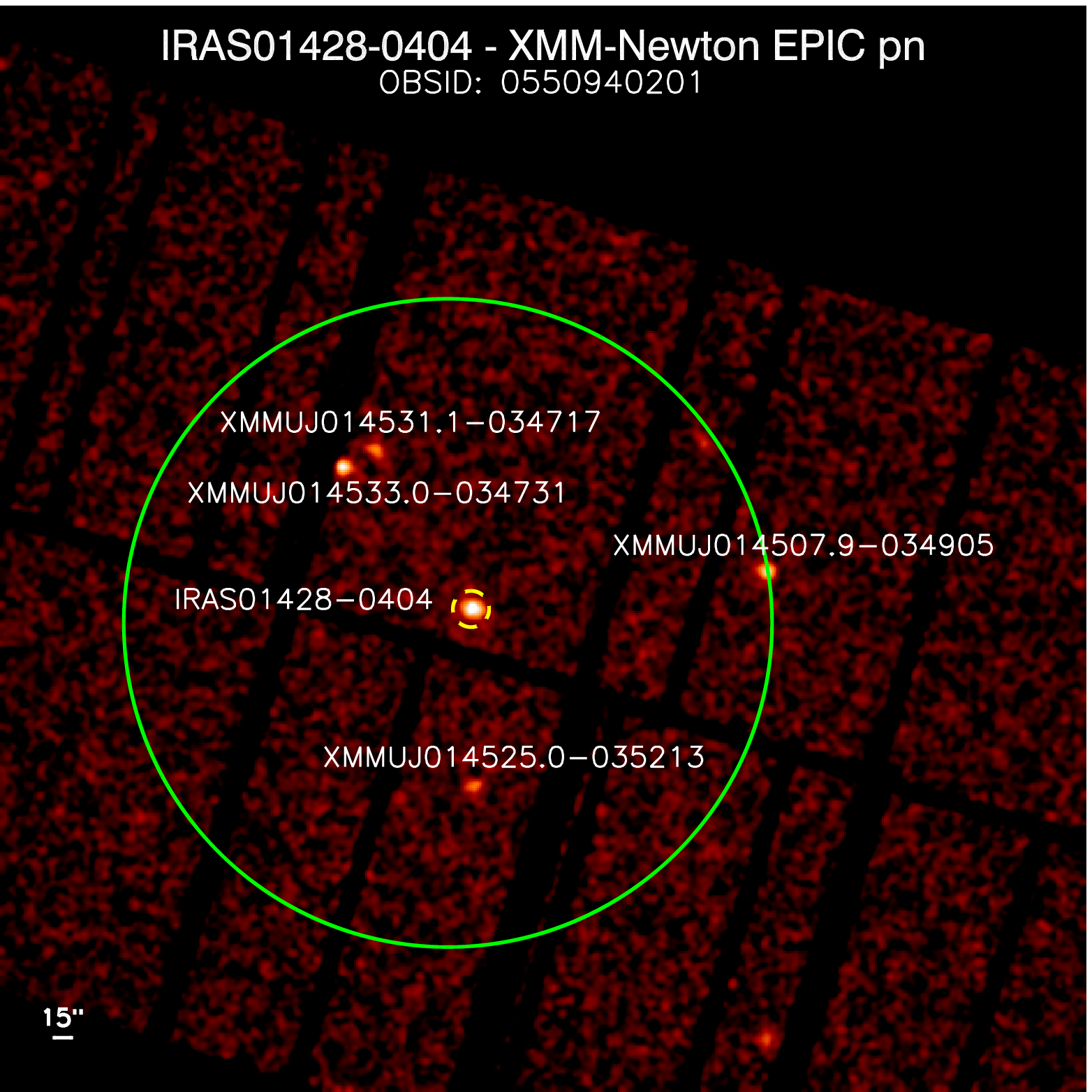,width=0.45\textwidth}
\end{center}
\caption{\label{irassed}\textit{Left:} Spectral Energy Distribution of IRAS
01428-0404 and best fit model (black solid line). Red-dotted line is the AGN
component and green-dashed line is the SB component. See text for details. \textit{Right:}  XMM-\textit{Newton} EPIC pn field of IRAS 01428-0404.
The extraction region for the pn spectrum is shown in yellow. The large green
circle is the extraction region for the \textit{ASCA} SIS1 spectrum, which
includes several other bright sources (see Table~\ref{iras_sources}).}
\end{figure*}

The only evidence potentially at odds with the Compton-thick interpretation of
IRAS~01428-0404 is represented by the larger X-ray fluxes measured in the past \textit{ASCA} observation.
Therefore, we downloaded the \textit{ASCA} data from the Tartarus
database\footnote{http://tartarus.gsfc.nasa.gov}, and re-analysed the GIS and
SIS spectra. The 0.5-2 keV energy band can be fitted with a very steep powerlaw
absorbed by the Galactic column density, for a flux in the same band of
($3.0\pm0.3$)$\times$10$^{-13}$ erg s$^{-1}$ cm$^{-2}$. In the 2-10 keV band, the
source is undetected ($3\sigma$ confidence level) by the SIS0 and the GIS2,
while it is formally detected ($\simeq4\sigma$) by the SIS1 and the GIS3. When
fitted with a pure reflection component, we recover a 2-10 keV flux of
($2.3\pm1.0$)$\times$10$^{-13}$ erg s$^{-1}$ cm$^{-2}$. Both the soft and the
hard X-ray flux are, therefore, roughly a factor of 10 larger than in our
XMM-\textit{Newton} observation. However, in the \textit{ASCA} source extraction
region there are several other sources, four of them with a comparable X-ray
flux to the target (see right panel of Fig.~\ref{irassed}). None of these
sources have been identified yet. Their X-ray spectra can be fitted with a
simple power law, absorbed by the Galactic column density. We give basic
information and X-ray fluxes in Table~\ref{iras_sources} for each of them. The
combined fluxes of these sources and IRAS~01428-0404 is consistent with the
\textit{ASCA} observed flux in the 2-10 keV band, while it is still a factor of
$~3$ lower in the 0.5-2 keV band. However, there is no reason to assume that the
varying source is IRAS 01428-0404.

\begin{table*}
\caption{\label{iras_sources}List of unidentified X-ray sources detected in the
XMM-\textit{Newton} observation of IRAS~01428-0404, within the source extraction
region adopted for the \textit{ASCA} observation. See right panel of
Fig~\ref{irassed} and text for details.}
\begin{center}
\begin{tabular}{llllll}
\textbf{Source} & \textbf{RA} & \textbf{DEC} & $\mathbf\Gamma$ & $\mathbf{F_s}$
& $\mathbf{F_h}$ \\
(1) & (2) & (3) & (4) & (5) & (6)\\
\hline
XMMU J014507.9-034905 & 01:45:07.9 & -03:49:05.8 & $1.6^{+0.2}_{-0.3}$ &
$2.5\pm0.4$ & $6.1\pm0.7$\\
XMMU J014525.0-035213 & 01:45:25.0 & -03:52:13.0 & $1.8^{+0.6}_{-0.5}$ &
$1.0\pm0.3$ & $1.8\pm0.5$\\
XMMU J014531.1-034717 & 01:45:31.1 & -03:47:17.0 & $2.1\pm0.6$ & $1.1\pm0.3$ &
$1.1\pm0.2$\\
XMMU J014533.0-034731 & 01:45:33.0 & -03:47:31.4 & $1.6^{+0.4}_{-0.3}$ &
$1.9\pm0.4$ & $4.8\pm1.9$\\
\hline
\end{tabular}
\end{center}
\textsc{Notes.}-- Col. (1) Name of the source following the naming conventions
suggested by the XMM-\textit{Newton} team. (2), (3) RA and DEC (Equatorial
J2000) (4) X-ray photon index (5), (6) 0.5-2 and 2-10 keV fluxes are in
10$^{-14}$ erg s$^{-1}$ cm$^{-2}$.
\end{table*}

\subsection{\label{mrk273x}Mrk273x}

Mrk~273x was classified as a Seyfert 2 galaxy by \citet{xia99} on the basis of
the large FWHM of the $\mathrm{[{O\,\textsc{iii}}]}$ emission line and the ratio
$\mathrm{[{O\,\textsc{iii}}]}5007/\mathrm{H\beta}>3$. Unfortunately, during our
optical observation the source was observed very close to morning twilight with
very large fringing and some sort of electronic noise, so the spectrum is rather
less than optimal, making it difficult to exclude the presence of a broad
component of the H$\beta$. Moreover, the redshift of the source (0.458) only
allowed us to observe the $\mathrm{[{O\,\textsc{iii}}]}5007$ and H$\beta$
wavelength range, thus limiting our diagnostics (Fig.~\ref{mrk273x_tng}).
Therefore, we can basically confirm the results presented by \citet{xia99},
without shedding more details on the optical classification of the source and the
presence of broad components of the permitted lines (see Tables~\ref{tableopt}
and \ref{mrk273x_opt}). A spectrum encompassing the H$\alpha$ emission line
region is needed to settle the issue.

\begin{table}
\caption{\label{mrk273x_opt}Mrk273x: optical emission lines in the \textit{TNG}
LRS spectrum (see Fig.~\ref{mrk273x_tng}).}
\begin{center}
% use packages: array
\begin{tabular}{lllll}
Line & $\lambda_\mathrm{l}$ & FWHM & Flux \\
(1) & (2) & (3) & (4)\\
\hline
 &  &  & \\
H$\beta$ & 4861.33 & $800\pm80$ & $3.8\pm1.9$ \\[1ex]
$\mathrm{[{O\,\textsc{iii}}]}$ & 4958.92 & $800\pm80$ &
$8\pm2$\\[1ex]
$\mathrm{[{O\,\textsc{iii}}]}$ &  5006.85 & $800\pm80$ &
$27\pm3$\\[1ex]
 &  &  &  \\
\hline
\end{tabular}
\end{center}
\textsc{Notes.}-- Col. (1) Identification. Col (2) Laboratory wavelength (\AA),
in air \citep{bowen60}. Col. (3) km s$^{-1}$ (instrumental resolution not removed).
 Col. (4) $10^{-16}$ erg cm$^{-2}$ s$^{-1}$.

\end{table}

\begin{figure}
\begin{center}
\epsfig{file=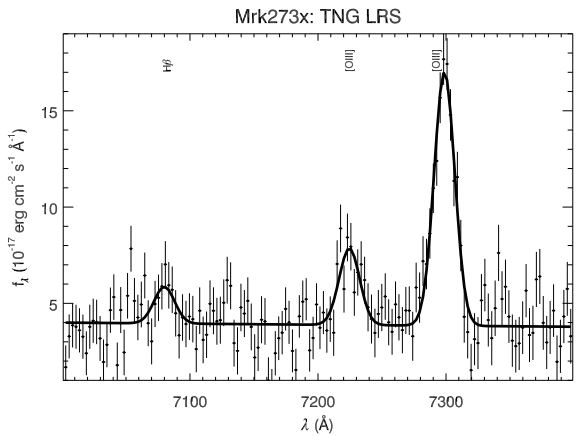,width=0.45\textwidth}
\end{center}
\caption{\label{mrk273x_tng}Mrk273x: TNG spectrum and best fit (see Table~\ref{mrk273x_opt}).}
\end{figure}

In X-rays, \textit{ASCA} and \textit{BeppoSAX} could not spatially resolve
Mrk273x from the neighbour Mrk273 \citep{iwa99,ris00}. Subsequent observations
with \textit{Chandra}, XMM-\textit{Newton} and \textit{Suzaku} provided an
unabsorbed powerlaw spectrum, with a large luminosity ($\simeq10^{44}$ erg
s$^{-1}$) and small variability on short and long timescales
\citep{xia02,bal05,teng09}. Our XMM-\textit{Newton} observation was divided into
four distinct exposures, all plagued by very high particle background. Even after the optimization method described in
Sect.~\ref{xmmred}, the resulting spectrum is very poor (see Fig.~\ref{mrk273x_xmm}), and can be fitted by a simple powerlaw
($\Gamma=1.6^{+0.5}_{-0.4}$) absorbed by the Galactic column density
($\chi^2=17/20$ d.o.f.). Any absorption in excess, at the redshift of the
source, can be constrained to be lower than $2\times10^{21}$ cm$^{-2}$. A more
stringent upper limit ($4.5\times10^{20}$ cm$^{-2}$) was found by \citet{bal05}
with the previous XMM-\textit{Newton} observation. The 2-10 (0.5-2) keV flux is
$1.1\pm0.6$ ($0.45\pm0.09$) $\times10^{-13}$ erg cm$^{-2}$ s$^{-1}$,
corresponding to a 2-10 (0.5-2) keV luminosity of $7.3\pm2.9$ ($3.0\pm0.6$) $\times10^{43}$ erg s$^{-1}$.
The 2-10 keV to $\mathrm{[{O\,\textsc{iii}}]}$ luminosity ratio is, therefore,
$\log{\frac{L_X}{L_{[OIII]}}}\simeq1.6$, which is perfectly consistent with the
source being unobscured \citep[see e.g.][]{lamastra09}, even considering the
(unknown, due to the lack of the H$\alpha$ emission line in the optical
spectrum) Balmer decrement to be applied to correct for reddening the
$\mathrm{[{O\,\textsc{iii}}]}$ luminosity.

\begin{figure}
\begin{center}
\epsfig{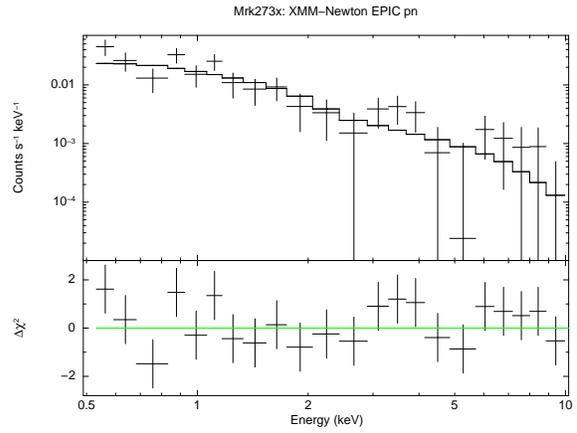}
\end{center}
\caption{\label{mrk273x_xmm}Mrk273x: the XMM-\textit{Newton} observation. EPIC pn spectrum, along with the best fit and residuals in terms of
$\Delta\chi^2$. See text for details.}
\end{figure}

\subsection{NGC3147}

The classification of NGC~3147 as a Seyfert 2 was originally given by \citet{hfs97}. \textit{ASCA} provided the first X-ray spectrum, without significant absorption, as later confirmed by \textit{BeppoSAX} \citep{dad07} and \textit{Chandra}, which also showed that no off-nuclear source can significantly contribute to the nuclear emission \citep{tw03}. The simultaneous X-rays and optical observations of NGC~3147 were analysed in detail by \citet{bianchi08a}, who confirmed the lack of broad permitted lines in the optical spectrum and of X-ray absorption, strongly suggesting that the source is a ``true type'' Seyfert 2. These conclusions were further refined thanks to a long broad-band \textit{Suzaku} observation, which allows only for a peculiar Compton-thick source dominated by an highly ionised and compact reflector as a viable alternative \citep{matt12}. However, this solution is strongly disfavoured by the observed X-ray variability on yearly time scales of the source. That the variation cannot be
due to a confusing source is demonstrated by the fact that the highest flux has been measured by the best spatial resolution satellite \citep[e.g.][and references therein]{matt12}. This again leaves the ``true'' Seyfert 2 nature of NGC 3147 as the most likely explanation.

\subsection{\label{3660}NGC3660}

NGC~3660 was optically classified as a Seyfert 2 by \citet{moran96} and \citet{gu06}, but as a composite/transition Seyfert 2/starburst galaxy by
\citet{kol83}, \citet{ccd98} and \citet{gonc99}. This could be due to the presence of a very compact nuclear starburst detected through the 3.3 $\mu$m polycyclic aromatic hydrocarbon (PAH) emission feature \citep{ima03}. A weak broad component of the H$\alpha$ and/or the H$\beta$ line was reported by \citet{kol83}, \citet{gonc99} and \citet{cidfernandes04}, after removal of the starlight.
We show in Fig.~\ref{ngc3660_not} our new optical spectra: the optical lines ratios are more typical of a Seyfert galaxy, but we confirm the possible contamination by a starburst region, mostly in the NOT spectrum (see Table~\ref{tableopt} and Fig.~\ref{diagn}). Both spectra also confirm the possible presence of a broad (FWHM$\simeq3000$ km s$^{-1}$) component of the H$\alpha$ emission line, but  with a flux at most 0.7 (0.8) that of the narrow core, in the NOT (TNG) spectrum. Fitting a component with the same width at the H$\beta$ wavelength results in an upper limit to the flux of only 0.1 (0.5) with respect to the narrow core.

\begin{figure*}
\begin{center}
\epsfig{file=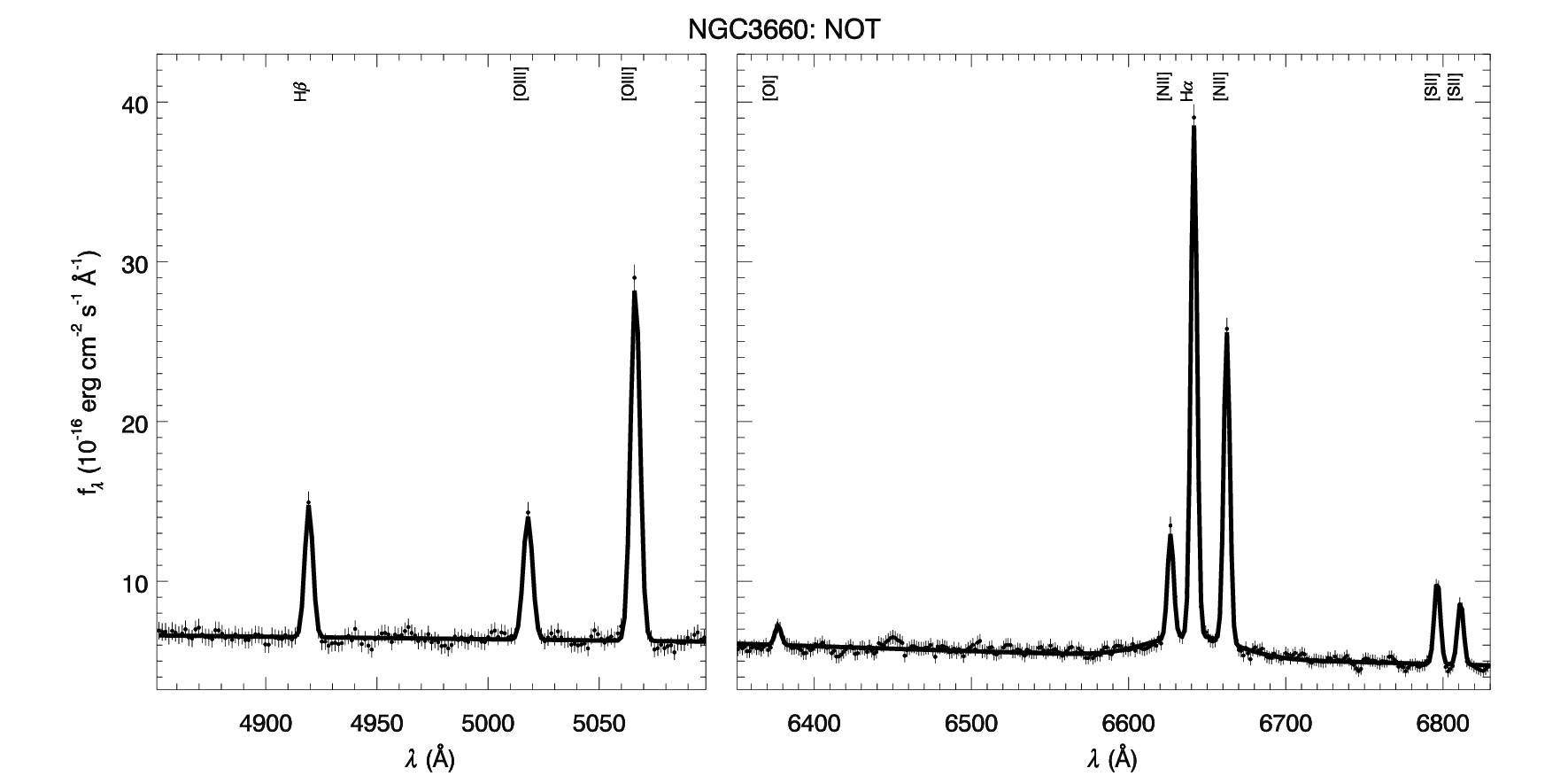,width=\textwidth}
\epsfig{file=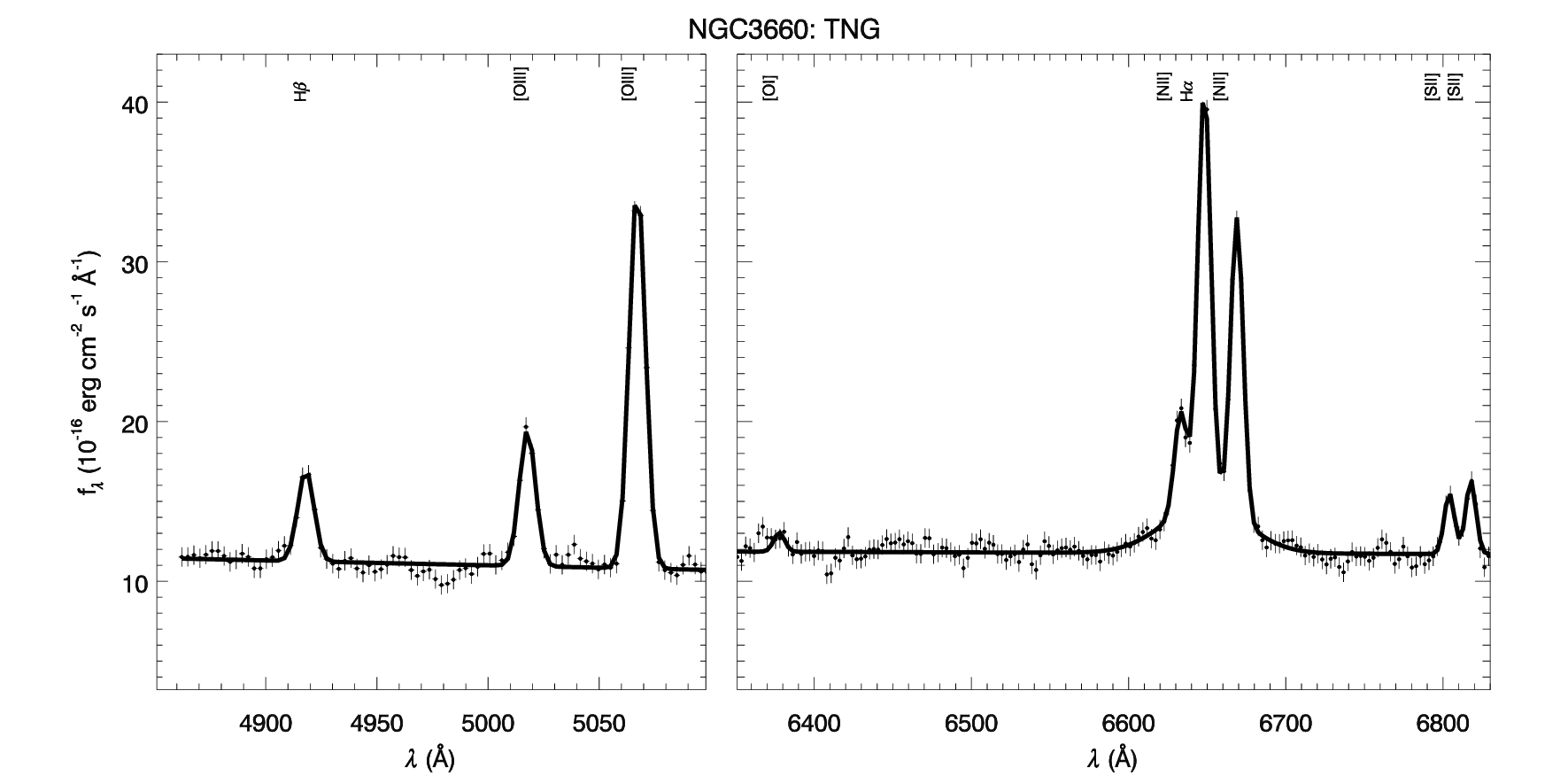,width=\textwidth}
\end{center}
\caption{\label{ngc3660_not}NGC~3660: NOT and TNG optical spectra along with best fits (see Table~\ref{ngc3660_opt} and \ref{ngc3660opt2}).}
\end{figure*}

\begin{table}
\caption{\label{ngc3660_opt}NGC3660: optical emission lines in the NOT
spectrum (see Fig.~\ref{ngc3660_not}).}
\begin{center}
% use packages: array
\begin{tabular}{lllll}
Line & $\lambda_\mathrm{l}$ & FWHM & Flux \\
(1) & (2) & (3) & (4) \\
\hline
 &  &  & \\
H$\beta$ & 4861.33 & $250\pm30$ & $3.8\pm0.4$ \\[1ex]
$\mathrm{[{O\,\textsc{iii}}]}$ & 4958.92 & $280\pm10$ &
$4.0\pm0.3$\\[1ex]
$\mathrm{[{O\,\textsc{iii}}]}$ & 5006.85 & $280\pm10$ &
$11.6\pm0.5$\\[1ex]
$\mathrm{[{O\,\textsc{i}}]}$ & 6300.32 & $240\pm110$ &
$0.5\pm0.2$\\[1ex]
$\mathrm{[{N\,\textsc{ii}}]}$ & 6548.06 & $200\pm10$ &
$3.17\pm0.13$\\[1ex]
H$\alpha$ & 6562.79 & $200\pm10$ & $15.5\pm0.5$ \\[1ex]
H$\alpha$ & 6562.79 & $2900\pm500$ & $9.5\pm1.4$ \\[1ex]
$\mathrm{[{N\,\textsc{ii}}]}$ & 6583.39 & $200\pm10$ &
$9.5\pm0.4$\\[1ex]
$\mathrm{[{S\,\textsc{ii}}]}$ & 6716.42 & $200\pm20$ &
$2.5\pm0.3$\\[1ex]
$\mathrm{[{S\,\textsc{ii}}]}$ & 6730.78 & $200\pm20$ &
$1.9\pm0.2$\\
 &  &  &  \\
\hline
\end{tabular}
\end{center}
\textsc{Notes.}-- Col. (1) Identification. Col (2) Laboratory wavelength (\AA),
in air \citep{bowen60}. Col. (3) km s$^{-1}$ (instrumental resolution not removed).
 Col. (3) $10^{-15}$ erg cm$^{-2}$ s$^{-1}$.

\end{table}

\begin{table}
\caption{\label{ngc3660opt2}NGC3660: optical emission lines in the TNG
spectrum (see Fig.~\ref{ngc3660_not}).}
\begin{center}
% use packages: array
\begin{tabular}{llll}
Line & $\lambda_\mathrm{l}$ & FWHM & Flux \\
(1) & (2) & (3) & (4) \\
\hline
 &  &  &  \\
H$\beta$ & 4861.33 & $520\pm100$ & $5.1\pm0.8$ \\[1ex]
$\mathrm{[{O\,\textsc{iii}}]}$ & 4958.92 & $510\pm20$ &
$7.7\pm0.6$\\[1ex]
$\mathrm{[{O\,\textsc{iii}}]}$ & 5006.85 & $510\pm20$ &
$21.6\pm0.7$\\[1ex]
$\mathrm{[{O\,\textsc{i}}]}$ & 6300.32 & $370^*$ & $1.0\pm0.6$\\[1ex]
$\mathrm{[{N\,\textsc{ii}}]}$ & 6548.06 & $420\pm20$ &
$6.2\pm0.2$\\[1ex]
H$\alpha$ & 6562.79 & $440\pm20$ & $27.0\pm0.8$ \\[1ex]
H$\alpha$ & 6562.79 & $2700\pm300$ & $18.7\pm2.5$ \\[1ex]
$\mathrm{[{N\,\textsc{ii}}]}$ & 6583.39 & $420\pm20$ &
$18.7\pm0.7$\\[1ex]
$\mathrm{[{S\,\textsc{ii}}]}$ & 6716.42 & $340\pm50$ &
$3.1\pm0.6$\\[1ex]
$\mathrm{[{S\,\textsc{ii}}]}$ & 6730.78 & $340\pm50$ &
$3.7\pm0.6$\\
 &  &  &  \\
\hline
\end{tabular}
\end{center}
\textsc{Notes.}-- Col. (1) Identification. Col (2) Laboratory wavelength (\AA),
in air \citep{bowen60}. Col. (3) km s$^{-1}$ (instrumental resolution not removed).
 Col. (4) $10^{-15}$ erg cm$^{-2}$ s$^{-1}$.

\end{table}

In order to confirm or reject the signatures of a (weak or heavily reddened) BLR in the optical spectrum, we also observed NGC~3660 in the near infrared, with two different grisms. There is no detection of any broad permitted hydrogen Paschen lines in either spectra (see Fig.~\ref{ngc3660_nics} and Table~\ref{ngc3660nir}). In particular, the tightest upper limit on a broad component of the Pa$\beta$ (once fixed the FWHM to the one possibly found in the optical spectra) comes from the spectrum with the highest resolution: its flux is at most 0.4 that of the narrow component. Therefore, we conclude that the BLR is not visible in the infrared, and the weak broad component possibly detected in the optical spectra is likely an artefact of a bad modelling of the continuum.

\begin{table}
\caption{\label{ngc3660nir}NGC3660: NIR emission lines in the TNG NICS spectrum with grism IJ (see Fig.~\ref{ngc3660_nics}).}
\begin{center}
% use packages: array
\begin{tabular}{llll}
Line & $\lambda_\mathrm{l}$ & FWHM & Flux \\
(1) & (2) & (3) & (4) \\
\hline
 &  &  &  \\
$\mathrm{[{S\,\textsc{iii}}]}$ & 9068.6 & $490\pm50$ & $12.7\pm2.3$ \\[1ex]
$\mathrm{[{S\,\textsc{iii}}]}$ & 9530.6 & $490\pm50$ & $28.0\pm2.4$\\[1ex]
He I & 10830.3 & $490\pm60$ & $22.0\pm2.2$\\[1ex]
Pa $\gamma$ & 10938.1 & $490\pm60$ & $3.8\pm1.7$\\[1ex]
$\mathrm{[{Fe\,\textsc{ii}}]}$ & 12566.8 & $360\pm60$ & $5.8\pm1.5$ \\[1ex]
Pa $\beta$ & 12818.1 & $360\pm60$ & $10.4\pm2.1$\\[1ex]
 &  &  &  \\
\hline
\end{tabular}
\end{center}
\textsc{Notes.}-- Col. (1) Identification. Col (2) Laboratory wavelength (\AA),
in air (NIST Atomic Spectra Database (ver. 4.1.0): http://physics.nist.gov/asd Col. (3) km s$^{-1}$ (instrumental resolution not removed).
 Col. (4) $10^{-15}$ erg cm$^{-2}$ s$^{-1}$.

\end{table}

\begin{figure*}
\begin{center}
\epsfig{file=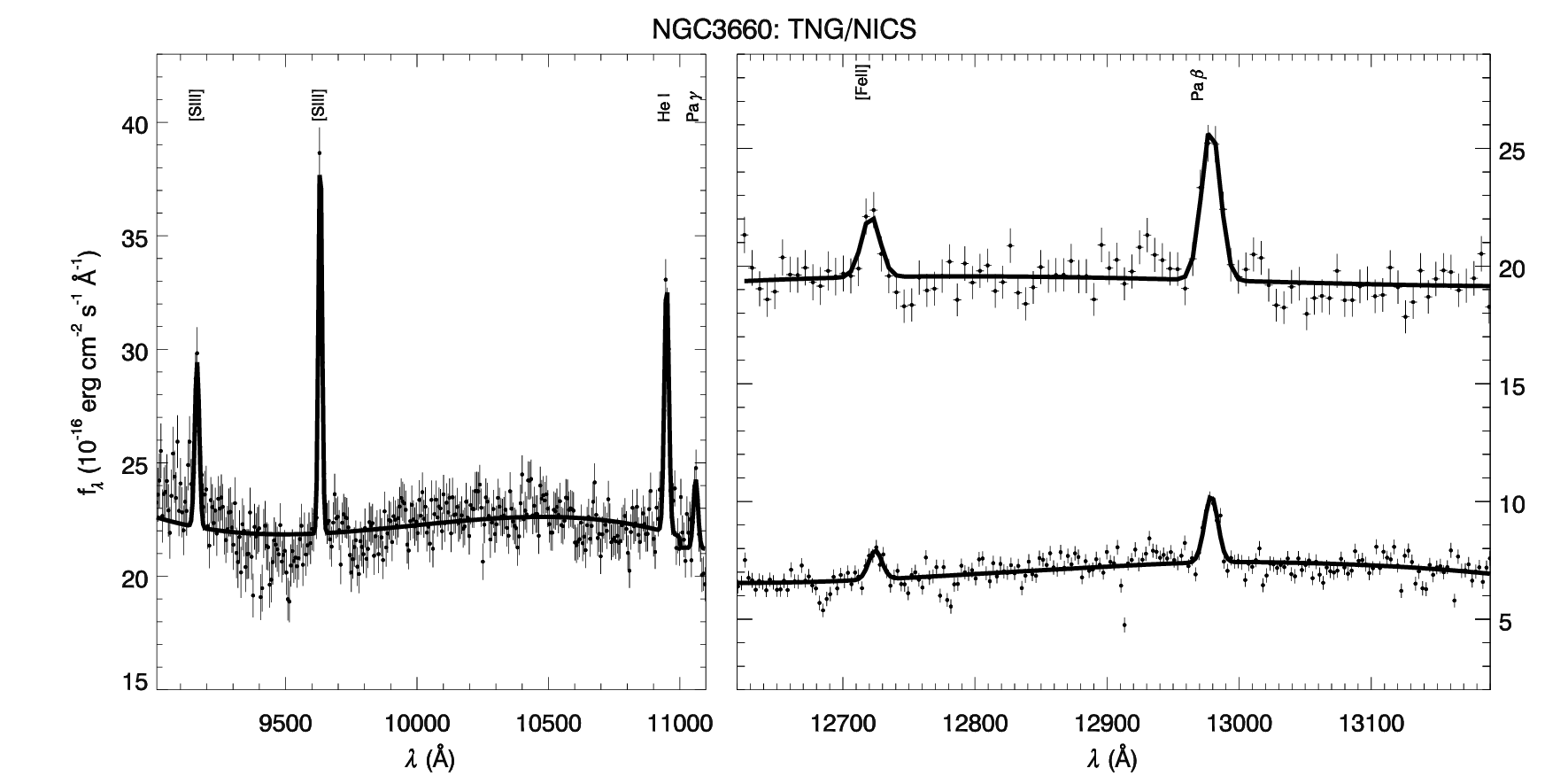,width=\textwidth}
\end{center}
\caption{\label{ngc3660_nics}NGC~3660: TNG/NICS spectra. In the left panel, the spectrum taken with grism IJ is shown, while in the right panel both the IJ (upper spectrum) and the JS (lower spectrum, rescaled for illustration purposes) are shown.}
\end{figure*}

In X-rays, NGC~3660 was detected in the \textit{ROSAT} All-Sky Survey with a 0.1-2.4 keV flux of $1.4\times10^{-12}$ erg s$^{-1}$ cm$^{-2}$ \citep{boller92}. The \textit{ASCA} spectrum is fitted
well by a power law with no absorption above the Galactic column, for a 2-10 keV flux of $2.3\times10^{-12}$ erg s$^{-1}$ cm$^{-2}$ \citep{brna08}. Interestingly, the same authors report a significant variability on short time-scales. The \textit{BeppoSAX} data confirm the absence of absorption in excess to that of the Galaxy, but the 20-100 keV flux measured by the PDS is well above what is predicted extrapolating the spectrum observed below 10 keV, suggesting a Compton-thick absorber along the line of sight, although at odds with the upper limit on the EW of the iron K$\alpha$ line of 230 eV \citep{dad07}. Moreover, the source is not detected by the \textit{Swift} BAT hard X-ray survey, even if the reported BeppoSAX PDS flux is larger than its flux limit \citep{cusu10}.

\begin{figure*}
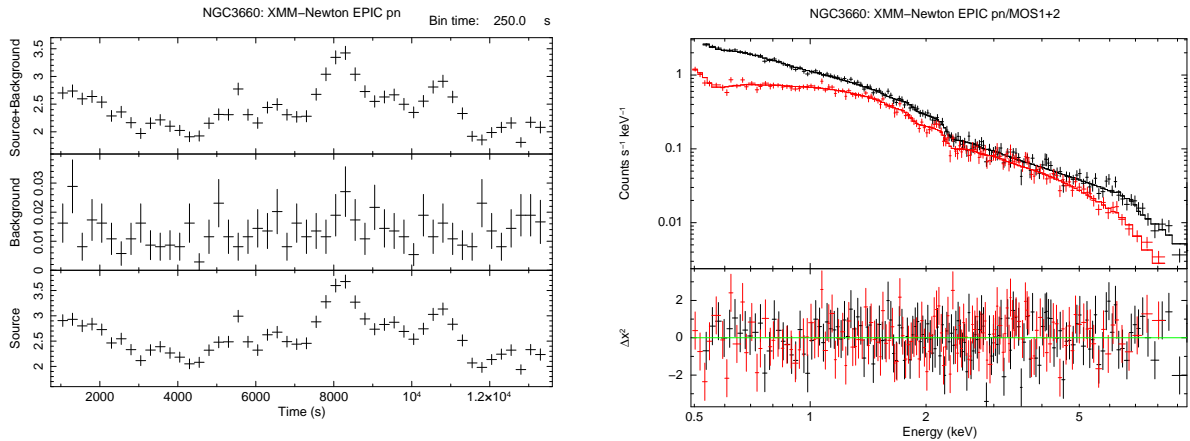

\begin{center}
\epsfig{file=ngc3660_pn_lc.ps,angle=-90,width=0.45\textwidth}
\epsfig{file=ngc3660_xmm.ps,angle=-90,width=0.45\textwidth}
\end{center}
\caption{\label{3660xmm}NGC~3660: the XMM-\textit{Newton} observation. \textit{Left:} EPIC pn lightcurve in the full band (0.3-10 keV). \textit{Right:} EPIC pn (black) and co-added MOS (red) spectra, along with the best fit and residuals in terms of $\Delta\chi^2$. See text for details.}
\end{figure*}

Our XMM-\textit{Newton} data confirm the rapid variability of NGC~3660 both in the soft and in the hard X-rays, with no strong hints for variations of the hardness ratio (see left panel of Fig~\ref{3660xmm}). The EPIC pn and co-added MOS spectra cannot be fitted by a simple power law absorbed by the Galactic column density ($\chi^2=535/281$ d.o.f.), mainly due to a clear excess below 0.8 keV. The latter can be modelled by a blackbody emission, at a temperature of $T=0.082\pm0.008$ keV, resulting in an acceptable fit ($\chi^2=310/279$ d.o.f.), with a power law index $\Gamma=1.99\pm0.03$. The addition of a neutral iron K$\alpha$ narrow ($\sigma=0$ eV) emission line at an energy fixed to 6.4 keV produces a marginal improvement of the fit ($\Delta\chi^2=4$ for one less degree of freedom}. Its flux is $\left(1.8\pm1.5\right)\times10^{-6}$ ph cm$^{-2}$ s$^{-1}$, and the EW=$60\pm50$ eV. A further improvement is achieved by adding a Compton reflection component (modelled with \textsc{pexrav}, with the photon index 
tied to that of the primary powerlaw, the inclination angle fixed to $30\degr$ and no cutoff energy), without changing significantly the other parameters, with the exception of the iron line, whose flux is now an upper limit ($<2.3\times10^{-6}$), and a rather unconstrained Compton reflection fraction ($R=2.4\pm1.1$). The fit is now perfectly acceptable ($\chi^2=291/277$ d.o.f.) and is shown in the right panel of Fig~\ref{3660xmm}. The 0.5-2 keV flux is $\left(2.45\pm0.05\right)\times10^{-12}$ erg s$^{-1}$ cm$^{-2}$, while the 2-10 keV flux is $\left(3.09\pm0.05\right)\times10^{-12}$ erg s$^{-1}$ cm$^{-2}$. They correspond to luminosities of $\left(9.5\pm0.2\right)\times10^{41}$ erg s$^{-1}$ and $\left(1.03\pm0.02\right)\times10^{42}$ erg s$^{-1}$, respectively.

\subsection{NGC4698}

\begin{figure*}
\begin{center}
\epsfig{file=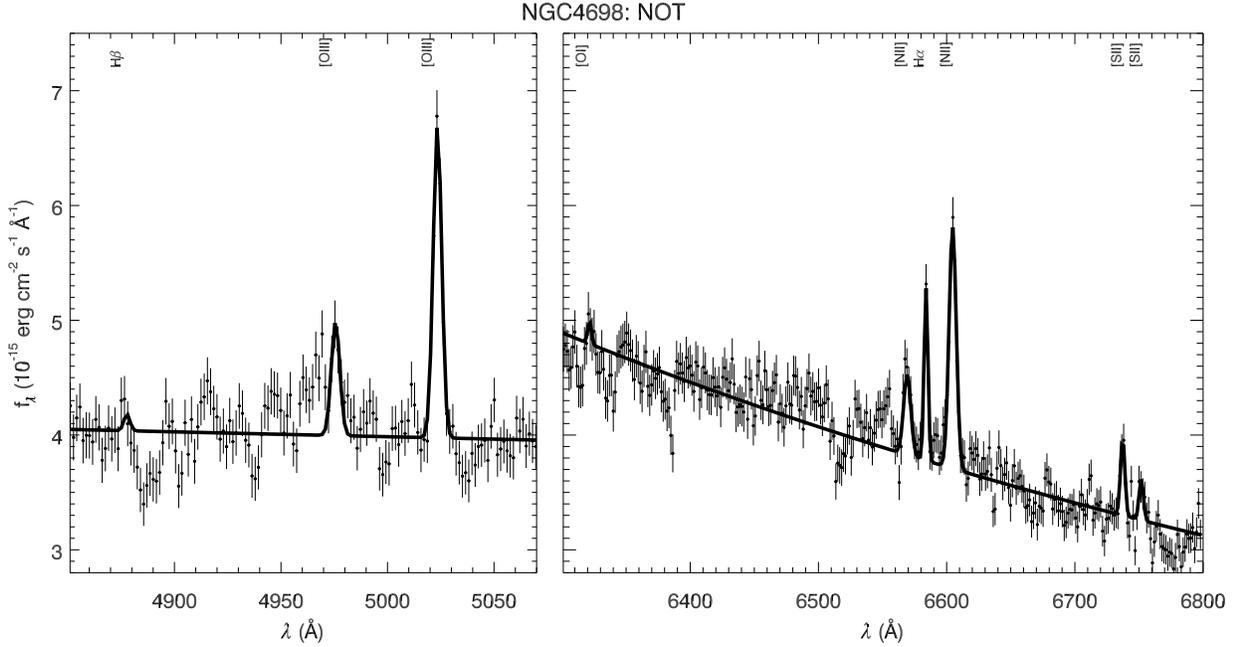,width=\textwidth}
\end{center}
\caption{\label{4698_optical}NGC~4698: NOT optical spectrum along with best fit (see Table~\ref{ngc4698opt}).}
\end{figure*}

NGC~4698 was classified as a Seyfert 2, with no trace of broad H$\alpha$, by \citet{hfs97} and \citet{ho97}. Our optical spectrum, shown in Fig.~\ref{4698_optical}, confirms unambiguously this classification (see Table~\ref{ngc4698opt} and Fig.~\ref{diagn}).

\begin{table}
\caption{\label{ngc4698opt}NGC4698: optical emission lines in the NOT spectrum (see Fig.~\ref{4698_optical}).}
\begin{center}
% use packages: array
\begin{tabular}{llll}
Line & $\lambda_\mathrm{l}$  & FWHM & Flux \\
(1) & (2) & (3) & (4) \\
\hline
 &  &  &  \\
H$\beta$ & 4861.33 & $270^*$ & $<1.6$ \\[1ex]
$\mathrm{[{O\,\textsc{iii}}]}$ & 4958.92 & $270\pm30$ & $4.9\pm1.2$\\[1ex]
$\mathrm{[{O\,\textsc{iii}}]}$ & 5006.85 & $270\pm30$ & $13.5\pm1.4$ \\[1ex]
$\mathrm{[{O\,\textsc{i}}]}$ & 6300.32 & $150^*$ & $<1.6$ \\[1ex]
$\mathrm{[{N\,\textsc{ii}}]}$ & 6548.06 & $300\pm30$ & $4.9\pm0.4$ \\[1ex]
H$\alpha$ & 6562.79 & $150\pm30$ & $5.6\pm1.0$ \\[1ex]
$\mathrm{[{N\,\textsc{ii}}]}$ & 6583.39 & $300\pm30$ & $14.8\pm1.3$\\[1ex]
$\mathrm{[{S\,\textsc{ii}}]}$ & 6716.42 & $170\pm50$ & $2.9\pm0.8$\\[1ex]
$\mathrm{[{S\,\textsc{ii}}]}$ & 6730.78 & $170\pm50$ & $1.5\pm0.7$\\
 &  &  &  \\
\hline
\end{tabular}
\end{center}
\textsc{Notes.}-- Col. (1) Identification. Col (2) Laboratory wavelength (\AA),
in air \citep{bowen60}. Col. (3) km s$^{-1}$ (instrumental resolution not removed).
 Col. (4) $10^{-15}$ erg cm$^{-2}$ s$^{-1}$.

\end{table}

In X-rays, after the detection by \textit{Einstein} \citep{fabb92}, NGC~4698 was observed with \textit{ASCA}, with a reported 2-10 keV luminosity of $2.2\times10^{40}$ erg s$^{-1}$, and an upper limit of the order of $10^{21}$ cm$^{-2}$ to any neutral absorbing column density \citep{pappa01}. The high spatial resolution of \textit{Chandra} showed that the \textit{ASCA} spectrum was dominated by two nearby AGN, while the nuclear source has a significantly lower luminosity of $10^{39}$ erg s$^{-1}$ \citep[0.3-8 keV: ][]{gz03}. However, the nuclear spectrum is still unabsorbed ($N_H\simeq5\times10^{20}$ cm$^{-2}$). These results were confirmed by the XMM-\textit{Newton} observation, even if the source extraction region was partly contaminated by some off-nuclear sources in the host galaxy \citep{cappi06,gonz09}.

\begin{figure*}
\begin{center}
\epsfig{file=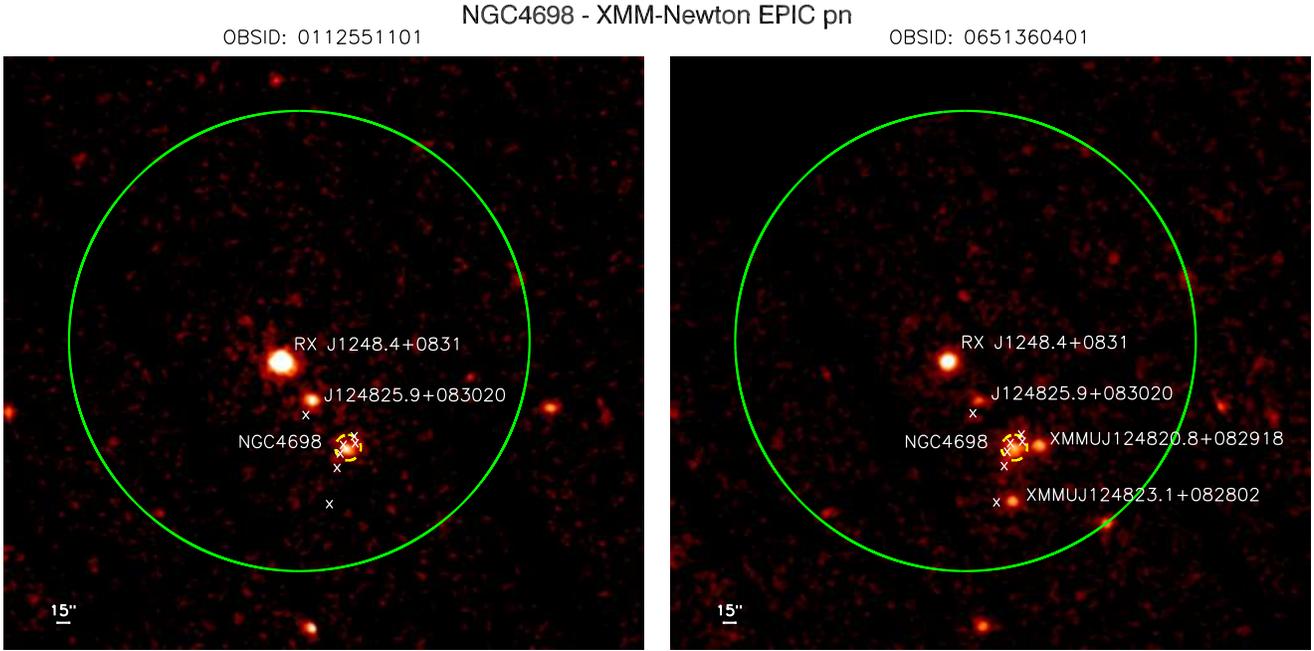,width=\textwidth}
\end{center}
\caption{\label{4698field}NGC~4698: XMM-\textit{Newton} EPIC pn fields for the two observations. The extraction regions for the pn spectra are shown in yellow.
The large green circle is the extraction region for the \textit{ASCA} SIS0 spectrum. White crosses are the sources detected in the \textit{Chandra}
observation, apart from NGC~4698, RX J1248.4+0831, and J124825.9+083020 \citep{gz03}. The two ULXs appeared in the second observation (see text and Table~\ref{ulx} for details) are marked in the right panel.}
\end{figure*}

\begin{figure}
\begin{center}
\epsfig{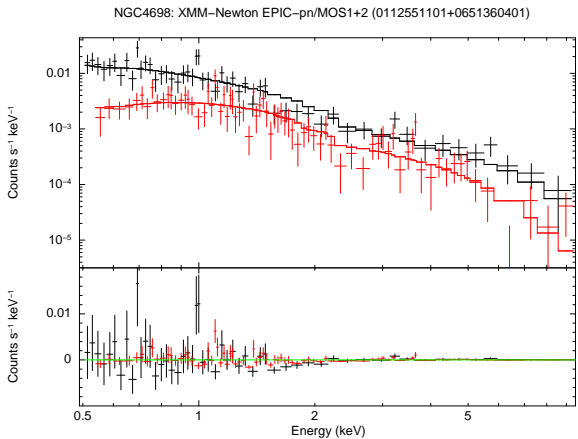}
\end{center}
\caption{\label{ngc4698_xmm}NGC~4698: EPIC pn (black) and co-added MOS (red) spectra, along with the best fit and residuals, for the combined XMM-\textit{Newton} observations. See text for details.}
\end{figure}

We show in Fig.~\ref{4698field} the EPIC pn fields of the old and our XMM-\textit{Newton} observations. As already shown by past works, the \textit{ASCA} extraction region is significantly contaminated, while the XMM-\textit{Newton} region that we chose contains $\simeq3$ off-nuclear sources, whose total X-ray flux contribute to $\simeq40\%$ of the observed flux in that region, assuming the fluxes reported by \citet{gz03}. After having verified that there is no significant variability during the observations and the spectra extracted from that region are consistent between the two observations, we decided to co-add them. The resulting X-ray spectrum (shown in Fig.\ref{ngc4698_xmm}) is well fitted by a powerlaw ($\Gamma=1.90\pm0.10$) absorbed by the Galactic column density (normalized Cash=146/126 d.o.f.). Only upper limits can be recovered for a local neutral absorbing column density ($N_H<1.9\times10^{20}$) and an iron K$\alpha$ emission line (EW$<450$ eV). The 0.5-2 keV flux is $\left(2.1\pm0.4\right)\times10^{-14}$ erg s$^{-1}$ cm$^{-2}$, while the 2-10 keV flux is $\left(3.0\pm0.5\right)\times10^{-14}$ erg s$^{-1}$ cm$^{-2}$. They correspond to luminosities of $\left(5.3\pm0.9\right)\times10^{38}$ and $\left(7.5\pm1.3\right)\times10^{38}$ erg s$^{-1}$, respectively. The 2-10 keV to $\mathrm{[{O\,\textsc{iii}}]}$ luminosity ratio is $\log{\frac{L_X}{L_{[OIII]}}}<0.15$, which is significantly lower than the average value for Compton-thin sources \citep[see e.g.][]{lamastra09}. Moreover, this ratio is an upper limit for two reasons: first, because the H$\beta$ is not detected in our optical spectrum, so we only have a lower limit of the Balmer decrement and, therefore, to the intrinsic [{O\,\textsc{iii}}] flux; moreover, the X-ray flux is an upper limit to the nuclear flux, which is likely to be contaminated by the off-nuclear sources detected by \textit{Chandra}. We can derive a better estimate of the ratio by using the \textit{Chandra} X-ray flux \citep[taken from][but rescaled to 2-10 keV]{gz03}, 
and
the [{O\,\textsc{iii}}] flux and Balmer decrement measured by \citet{ho97}. The resulting ratio is $\log{\frac{L_X}{L_{[OIII]}}}=-0.27$, suggesting that the source is Compton-thick \citep[see e.g.][]{lamastra09,mar12}. This is in agreement with the low $\mathrm{L_X/L_{12\,\mu m}}$ ratio, as reported by \citet{shi10}.

\subsubsection{The off-nuclear sources in the NGC~4698 field}

As a final note, it is interesting to mention the large variability of the off-nuclear sources between the two XMM-\textit{Newton} observations, as clearly shown in Fig.~\ref{4698field}. In particular, the two sources inside the $D_{25}$ of the galaxy that we label XMMU J124820.8+082918 and XMMU J124823.1+082802 are detected at high significance in the second observation, but not in the previous one. The fit with a simple powerlaw with Galactic intervening absorption is acceptable for both objects ($\chi^2=30/31$ d.o.f. and $32/26$ d.o.f., respectively), while more complicated spectral shapes are not required (see Table~\ref{ulx}). The luminosities of the two sources ($\simeq10^{39}$ erg s$^{-1}$ if at the distance of NGC~4698) put them in the Ultra-Luminous X-ray sources (ULX) regime, even if at the lower $L_X$ range. We can interpret these sources as variable X-ray binaries with a BH mass of the order of $\simeq10\,M_{\sun}$.

In the first XMM-\textit{Newton} observation, at the positions of these two sources we find only upper limits to the luminosities of $\simeq5$ and $\simeq3\times10^{38}$ erg s$^{-1}$, respectively. XMMU J124820.8+082918 is marginally detected in the \textit{Chandra} observation (performed one year later, in 2002) with a luminosity of $\simeq3\times10^{37}$ erg s$^{-1}$ \citep[estimated by re-scaling its net counts with respect to the net counts in the nucleus as reported in Table 1 of][]{gz03}. On the other hand, XMMU J124823.1+082802 is not detected even in the \textit{Chandra} observation, and might be a `new' source. It is indeed consistent with a possible supernova, both on energetics and spectral shape, since it can be fitted by an \textsc{apec} thermal component \citep[as for SN IIn for instance, see e.g][]{il03}, although the fit cannot be statistically preferred to a simple powerlaw.

No apparent optical counterpart is present on the available optical/IR/UV material, however none is from the epoch of the second XMM-Newton observation. ULXs are massive and therefore have young progenitors. It follows that the brightening or appearance of new sources is at odds with the optical classification of the nucleus of NGC~4698 with a population of age T $>$ 5 Gyr \citep{cor12}.  We might speculate that the merging responsible for the nuclear appearance still has an impact on the outskirts of the galaxy, where star formation is visible in two UV rings from \textit{GALEX} images \citep{ch09} at a rate of $\sim 0.1$ M$_{\sun}$/yr (I. De Looze, private communication), or that other mechanisms are at work.

\begin{table*}
\caption{\label{ulx}Properties of the two ULX candidates appeared in the new XMM-\textit{Newton} observation of NGC~4698. See text for details.}
\begin{center}
\begin{tabular}{lrrrrrrr} \hline
Name & RA & DEC & $\Gamma$ & $F_s$ & $F_h$ &  $L$ \\
(1) & (2) & (3) & (4) & (5) & (6) & (7) \\
\hline
XMMU J124820.8+082918 & 12:48:20.8 & +08:29:18.4 & $1.52\pm0.13$ & 1.91 & 4.6 & 1.6 \\
XMMU J124823.1+082802 & 12:48:23.1 & +08:28:02.5 & $1.56\pm0.16$ & 1.29 & 2.9 & 1.0 \\
\hline
\end{tabular}
\end{center}

\textsc{Notes.}-- Col. (1) Name of the source following the naming conventions suggested by the XMM-\textit{Newton} team. (2), (3) RA and DEC (Equatorial J2000) (4) X-ray photon index (5), (6) 0.5-2 and 2-10 keV fluxes are in 10$^{-14}$ erg s$^{-1}$ cm$^{-2}$, (7) 0.5-10 keV luminosities are in 10$^{39}$ erg s$^{-1}$
\end{table*}

\subsection{Q2130-431 and Q2131-427}

Q2130-431 and Q2131-427 belong to the `naked' AGN class, where the absence of broad emission lines is accompanied by strong optical variability, suggesting that the nucleus is seen directly \citep{hawk04}. The absence of significant absorption in the X-ray spectra was later confirmed by \textit{Chandra} snapshots \citep{gliozzi07}. The simultaneous X-rays and optical observations of Q2130-431 and Q2131-427 were analysed in detail by \citet{panessa09}. While the optical spectrum of Q2130-431 do show broad components of the Balmer lines, Q2131-427 appears to lack broad optical lines and X-ray absorption along the line of sight, thus being a true Type 2 Seyfert galaxy.

\section{Discussion}

\subsection{Rejected candidates: IC1631, IRAS01428-0404, NGC4698, Q2130-431}

The Seyfert 2 nature was confirmed by our new optical spectra for all the sources of our sample (but see Sect.~\ref{mrk273x} and \ref{mrk273x_disc} for the case of Mrk273x), with the exception of IC~1631 and Q2130-431. The classification of IC1631 was ambiguous in the literature, but all the line diagnostics derived from our optical spectrum clearly put the source in the region populated by starburst galaxies. The X-ray data are consistent with this interpretation, requiring a star formation rate of $\simeq10-15$ M$_\odot$/yr. Therefore, IC~1631 must be removed, once and for all, as a candidate unabsorbed Seyfert 2.

On the other hand, Q2130-431 clearly shows broad components of the Balmer lines in its optical spectrum. However, notwithstanding the negligible observed Balmer decrement on the broad lines (3.4), the flux of the H$\beta$ broad component appears quite weak with respect to the $\mathrm{[{O\,\textsc{iii}}]}$ line. However, the ratio between the luminosity of the broad H$\alpha$ line and the 2-10 keV luminosity ($\mathrm{L_{H\alpha}^{broad}/L_{2-10\,keV}}\simeq-1.7$) is still within the distribution presented by \citet{shi10} for type 1 and intermediate AGN. Therefore, the BLR in this source may be somewhat weaker than in `normal' AGN, but the source cannot be considered a true Type 2 Seyfert galaxy.

Other two sources, IRAS01428-0404 and NGC4698, are very likely Compton-thick. In both cases, the optical classification as a Seyfert 2 is unambiguous. On the other hand, the presence of a Compton-thick absorber cannot be directly confirmed by the X-ray spectral analysis, because of the weakness of the sources. However, the low 2-10 keV to $\mathrm{[{O\,\textsc{iii}}]}$ luminosity ratios are quite suggestive that the primary X-ray continuum is absorbed by a Compton-thick material along the line-of-sight, while we only observe a small fraction as reflected by circumnuclear matter. If this interpretation is correct, the two sources are standard Compton-thick Seyfert 2 galaxies, and must be cancelled from any list of unabsorbed Sy2 candidates.

\subsection{\label{mrk273x_disc}Mrk273x: a peculiar candidate}

One object of our initial sample, Mrk~273x, represents a good candidate as an unabsorbed Seyfert 2. However, the optical classification as a Seyfert 2 is only based on the lack of a broad component of the H$\beta$ emission line: a spectrum encompassing the H$\alpha$ emission line region is definitely needed to settle the issue. Interestingly enough, its X-ray luminosity is very large ($\simeq10^{44}$ erg s$^{-1}$). Adopting different bolometric corrections for the X-ray luminosity \citep{elvis94,mar04,vf09}, we obtain bolometric luminosities in the range $1-3\times10^{45}$  erg s$^{-1}$. Although no BH mass estimates are present in the literature, we can have a rough one by using the FWHM of the $\mathrm{[{O\,\textsc{iii}}]}$ emission line as a proxy of the stellar velocity dispersion \citep[see e.g.][]{gh05}, and then using the \citet{trem02} relation. With the FWHM reported by \citet{xia99}, we get a BH mass of $\simeq1\times10^8\,M_{\sun}$. Therefore, the Eddington rate of Mrk~273x would be of the order
$0.08-0.2$. If this source is confirmed to really lack the BLR, the models invoking a low accretion rate/low luminosity regime for its disappearance would be seriously challenged (see next section).

This source closely resembles 1ES 1927+654, a luminous type 2 Seyfert galaxy (L$_\mathrm{0.1-2.4 keV}$ $\sim$ 4.6 $\times$ 10$^{43}$ ergs sec$^{-1}$) which revealed persistent, rapid and large scale variations in ROSAT and \textit{Chandra} observations, and no X-ray absorption \citep{boll03}. Our campaign of quasi simultaneous TNG Optical/NIR and XMM-Newton observations have confirmed the `true' type 2 nature of this source, i.e., no broad optical emission line components nor X-ray absorption in excess to the Galactic one (Panessa et al. in preparation). However, the XMM-\textit{Newton} flux is ten times lower than the ROSAT one, so the engine might be still strongly active and variable (caught in a low flux state) or may be fading away. The Eddington ratio derived from the XMM-\textit{Newton} flux is L$_\mathrm{Bol}$/L$_\mathrm{Edd}$ $\sim$ 0.01, still at the borderline of the critical accretion rates predicted by theoretical models (see next section).

\subsection{\label{truetype2}True type 2 Seyfert 2s: NGC3147, NGC3660, and Q2131-427}

\begin{table}
\caption{\label{tt2properties}Main properties of the three true type 2 Seyfert 2s confirmed in this paper. See text for details.}
\begin{center}
\begin{tabular}{lccc} \hline
Name & L & $M_{BH}$ & $L_{bol}/L_{Edd}$\\
(1) & (2) & (3) & (4) \\
\hline
NGC3147 & $0.3-0.7$ & $20-62$ & $4\times10^{-5}-3\times10^{-4}$\\
NGC3660 & $1-2$ & $0.68-2.1$ & $4\times10^{-3}-2\times10^{-2}$ \\
Q2131-427 & $20-30$ & $77$ & $2-3\times10^{-3}$ \\
\hline
\end{tabular}
\end{center}

\textsc{Notes.}-- Col. (1) Name of the source. (2) Bolometric luminosity range in $10^{43}$ erg s$^{-1}$ (see text for details on the adopted methods for this estimate) (3) BH mass range in $10^7\,M_{\sun}$ (see text for the appropriate references) (4) Eddington ratio range
\end{table}

Out of the initial sample composed by 8 sources, three, namely NGC~3147, NGC~3660, and Q2131-427, do appear to simultaneously lack the broad optical lines and X-ray absorption along the line-of-sight\footnote{The optical and X-ray observations of NGC~3660 are not strictly simultaneous due to technical problems, but all the available optical spectra are consistent each other, as described in Sect.~\ref{3660}.}. In all cases, although a Compton-thick interpretation of the X-ray spectrum cannot be completely excluded \citep[see][]{matt12,brna08,panessa09}, the short-term X-ray variability of NGC3660, the variability on few-years timescale of NGC~3147, and the large optical brightness variability of Q2131-427 strongly support their unabsorbed nature. In order to rule out in these sources the presence of broad optical lines as for standard unobscured AGN, we should compare the weak broad components compatible with their optical spectra to the expectations for `normal' objects.

In the case of NGC~3147, \citet{bianchi08a} reported an upper limit to a broad (FWHM=2000 km s$^{-1}$) component of the H$\alpha$ corresponding to a luminosity of $2\times10^{38}$ erg s$^{-1}$. This value is significantly lower than expected from the X-ray luminosity, i.e. $\simeq5.6\times10^{40}$ erg s$^{-1}$, considering the relation presented by \citet{sl12}, based on a large sample of SDSS Type 1 AGN. Similarly, the observed luminosity of the weak broad component of the H$\alpha$ in NGC~3660 is $6.2\times10^{39}$ erg s$^{-1}$, again significantly lower than the expected value, $\simeq2.8\times10^{41}$ erg s$^{-1}$. For Q2131-427, the upper limit (FWHM=4000 km s$^{-1}$) is $8\times10^{40}$ erg s$^{-1}$ \citet{panessa09}, to be compared to the expected $8\times10^{42}$ erg s$^{-1}$. The same conclusions can be drawn by comparing the observed $\mathrm{L_{H\alpha}^{broad}/L_{2-10\,keV}}$ ratios of NGC~3147, NGC~3660, and Q2131-427 to the distribution presented by \citet{shi10} for type 1 and intermediate AGN.
 In the case of NGC~3147 this is also true for the $\mathrm{L_{H\beta}^{broad}/\nu L_\nu^{1.4\,GHz}}$ and $\mathrm{L_{H\alpha}^{broad}/L_\nu^{10\,\mu m}}$ ratios \citep{shi10}. Therefore, in these sources the emission of the BLR, if present, is much weaker than the one found in common Seyfert 1s, or the widths of the broad optical line components are much larger.

Reverberation mapping studies show that the radius of the BLR and the bolometric luminosity of the AGN always follow a tight L$^{1/2}$ relation \citep[see e.g.][who showed how the C~\textsc{iv} emission line lags follow this relation over more than 7 orders of magnitude in continuum luminosity]{kaspi07}. This relation is naturally explained by assuming that the outer boundary of the BLR is determined by the dust sublimation radius, which is set by the continuum luminosity. Once the BLR radius is determined by the luminosity, the BH mass sets the velocity of the gas at that distance. In formul\ae, combining Eq. 2 and 6 in \citet{sl12}, we get the expected FWHM of the broad optical lines as a function of the BH mass and the bolometric luminosity of the AGN:

\begin{equation}
\label{fwhmformula}
\mathrm{FWHM_{H_\alpha}} \simeq 7088 \, \left(\frac{M_{BH}}{10^8 M_{\sun}}  \right)^{0.49} \, \left(\frac{L_{bol}}{10^{44}} \right) ^{-0.26} \mathrm{km\,s^{-1}}
\end{equation}

The bolometric luminosity of NGC~3147 can be estimated from the 2-10 keV X-ray luminosity, whose average value around $3\times10^{41}$ erg s$^{-1}$ was measured by XMM-\textit{Newton} \citep{bianchi08a} and \textit{Suzaku} \citep{matt12}. Adopting different bolometric corrections for the X-ray luminosity \citep{elvis94,mar04,vf09}, we obtain bolometric luminosities in the range $3-7\times10^{42}$  erg s$^{-1}$. On the other hand, BH mass estimates in literature range from 2.0 to $6.2\times10^8\,M_{\sun}$ \citep{merl03,dd06}. Using these estimates in Eq.~\ref{fwhmformula}, the expected FWHM for the H$\alpha$ emission line produced in the BLR should be in the range $20\,000-40\,000$ km s$^{-1}$. Objects with such broad lines are expected to be extremely rare, if existing at all \citep[see e.g.][]{sl12}.

In the case of NGC~3660, from the X-ray luminosity measured in Sect.~\ref{3660} we get bolometric luminosities in the range $1-2\times10^{43}$ erg s$^{-1}$. With BH mass estimates ranging from $6.8\times10^6\,M_{\sun}$ \citep{mks10} to $2.1\times10^7\,M_{\sun}$ \citep{wz07}, the expected FWHM for the broad H$\alpha$ component is 2800-6000 km s$^{-1}$. This value is consistent with the FWHM of the broad component possibly detected in our optical spectra (but not confirmed in the infrared spectrum: see Sect.~\ref{3660}), whose luminosity, as shown above, is significantly lower than the one expected from a `normal' BLR.

Finally, from the X-ray luminosity of Q2131-427 and the BH mass ($7.7\times10^8\,M_{\sun}$) reported by \citet{panessa09}, we get bolometric luminosities in the range $2-3\times10^{44}$ erg s$^{-1}$, and an expected FWHM for the broad H$\alpha$ component of $14\,000-16\,000$ km s$^{-1}$ km s$^{-1}$. These values, even if less extreme than those derived for NGC~3147, are still exceptional.

The weakness of the broad optical lines may be due, as in normal Seyfert 2s, to dust extinction. The amount of dust required is very large, since the broad lines are not present even in the NIR spectra \citep[see Sect.~\ref{3660} for NGC~3660 and][for NGC~3147]{tran11}. However, in the X-ray spectra of these sources there are no signatures of absorption by cold gas along the line-of-sight, requiring an anomalous high fraction of dust associated to very little gas. Although deviations from the Galactic gas-to-dust ratios are very common in AGN \citep[e.g.][]{mai01}, there is always more gas than expected, likely explained by the presence of absorbing gas within the dust sublimation radius \citep[e.g.][and references therein]{bmr12}. The opposite situation, i.e. more dust than standard, would imply the presence of an unlikely physical mechanism able to suppress gas without destroying dust. In principle a very highly ionised dusty warm absorber could be difficult to detect in these X-ray spectra without high 
spectral resolution and high statistics. However, the properties of such a dusty warm absorber should be much more extreme than those typically found in Seyfert 1 galaxies \citep[e.g.][and references therein]{lee01}. Alternatively, an ad-hoc geometry could account for dust extinction of the BLR, but no gas absorption along the line-of-sight to the X-ray source. However, this peculiar geometry should also prevent us from seeing the optical broad lines even in polarized light: both sources have high-quality spectro-polarimetric data, with no evidence for an hidden BLR \citep{tran03,shi10,tran11}.

The most likely explanation for the absence of broad optical lines in these sources is that they intrinsically lack the BLR. The inability of some AGN to form the BLR is predicted by several theoretical models, all based on a rich literature elaborating the idea that the BLR is part of a disk wind \citep[e.g.][]{emm92,mur95,elvis00}. Evidence in favour of the presence of this wind in Seyfert galaxies is likely represented by the observation of X-ray and UV absorbers outflowing up to very high velocities \citep[e.g.][]{blust05,tomb10}. Both \citet{nic00} and \citet{trump11} assume that this disk wind originates at the radius where radiation pressure is equal to the gas pressure. If this radius becomes smaller than a characteristic critical radius, the disk wind cannot be launched, and the BLR cannot from. This critical radius can be identified with the innermost last stable orbit of a classic \citet{ss73} disk \citep{nic00}, or the transition radius to a radiatively inefficient accretion flow \citep{
trump11}. In both cases, the formation of the BLR is prevented for Eddington rates $\dot{m}=L_{bol}/L_{Edd}$ lower than a critical value, which can be expressed as $\dot{m}\simeq2.4\times10^{-3}\,M_8^{-1/8}$ and $\dot{m}\simeq1.3\times10^{-2}\,M_8^{-1/8}$, respectively, where $M_8$ is the BH mass in units of $10^8 M_{\sun}$. The Eddington rate of NGC~3147, with the same considerations as above, can be estimated in the range $4\times10^{-5}-3\times10^{-4}$, well below the thresholds predicted by both models. The same holds for Q2131-427, where $\dot{m}=2-3\times10^{-3}$. In the case of NGC~3660, $\dot{m}=4\times10^{-3}-2\times10^{-2}$, lower than the \citet{trump11} threshold, but only marginally consistent with the \citet{nic00} one in the low end.

\citet{trump11} support their model by showing an observational limit at $\dot{m}\simeq0.01$ between AGN with broad optical lines and (X-ray unobscured) AGN without. Another observational evidence of the existence of a minimum accretion rate for the formation of the BLR comes from several studies that point out the absence of broad optical lines in the spectra in polarized light of Seyfert 2s with low Eddington rates \citep[e.g.][]{nic03,bigu07,wu11,mar12}. In the recent analysis by \citet{mar12}, the threshold is found at $\dot{m}\simeq0.01$, which is in agreement both with the model and the data presented by \citet{trump11}.

Are true type Seyfert 2s rare objects? Apart from NGC~3147, NGC~3660, and Q2131-427, there are not many other strong representatives of this class in literature. If all true type Seyfert 2s are indeed low-accretors, their paucity should not be very surprising. As already noted by \citet{mar12} and \citet{bmr12}, when a sizeable sample of X-ray unobscured radio-quiet AGN with good-quality spectra is analysed \citep[e.g. CAIXA:][]{bianchi09}, only a few ($<5\%$) lie below the $\dot{m}\simeq0.01$ limit, with NGC~3147 and NGC~3660 among them. The fraction of low-accreting unabsorbed Seyfert 2 candidates rises up to 30\% in extensively studied samples derived from surveys \citep[COSMOS:][]{trump11}, but the lack of simultaneous optical and X-ray observations, and the low quality of the X-ray spectra, prevent us from drawing firm conclusions on their nature as genuine true type 2 AGN. Interestingly, a similar fraction of $\simeq25\%$ of low-accreting objects are found to lack the signatures of an hidden BLR in 
polarized light in obscured AGN \citep{mar12}. In this scenario, these sources would represent the obscured counterparts of true type Seyfert 2s, the only difference being the presence of an obscuring medium along the line of sight.

A separate discussion is probably needed for the so-called Low Luminosity AGN (LLAGN), which are mostly LINERs or transition objects, with bolometric luminosites significantly lower than $10^{42}$  erg s$^{-1}$ \citep[e.g.][]{ho97,thp00}. There are some LLAGN accreting at rates well below 0.01 with broad optical emission lines \citep[e.g., M81:][]{hfs96}. Recently, \citet{eh09} presented a comprehensive analysis of a sample of LLAGN with the aim to derive observationally where the separation between objects with BLR and those without lies. The separation they propose can be expressed in terms of accretion rate as $\dot{m}\simeq1.8\times10^{-6}\,M_8^{-1/3}$, much lower than the ones predicted by the \citet{trump11} and \citet{nic00} models (with which our results agree), despite adopting a similar disk-wind scenario. \citet{trump11} discussed in detail this discrepancy, without reaching a definite conclusion. It seems that the formation of the BLR in objects with very inefficient accretion regimes may be 
significantly different than what occurs in higher-luminosity AGN. Further studies on LLAGN are clearly fundamental to understand this issue and to shed light on the link between the accretion mechanisms and the formation of the BLR.

\section*{Acknowledgements}

We thank A. Laor and J. Stern for useful discussions, and the anonymous referee for helping us in improving the manuscript.
SB, GM, and AW acknowledge financial support from ASI (grant 1/023/05/0). FP acknowledges support from INTEGRAL ASI I/033/10/0, and FP and AW from ASI/INAF I/009/10/0. XB and FJC acknowledge partial financial support from the Spanish Ministerio de Ciencia e Innovaci\'on project AYA2010-21490-C02-01.

\bibliographystyle{mn2e}
\bibliography{sbs}

\label{lastpage}

\end{document}